\documentclass[aps,prd,10pt,twocolumn,groupedaddress,nofootinbib,floatfix]{revtex4-2}
\usepackage[T1]{fontenc}
\usepackage{amsmath,amssymb,bm}
\usepackage{graphicx,booktabs}
\usepackage{microtype}
\usepackage{xcolor}
\usepackage{hyperref}
\hypersetup{hidelinks,
 pdftitle={Energy-independent neutrino dephasing from an ultralight axion-like background},
 pdfauthor={B. A. Couto e Silva and B. L. Sanchez-Vega}}
\newcommand{\dd}{\mathrm{d}}
\newcommand{\ii}{\mathrm{i}}
\newcommand{\Tr}{\operatorname{Tr}}
\newcommand{\diag}{\operatorname{diatheg}}
\newcommand{\eV}{\,\mathrm{eV}}
\newcommand{\GeV}{\,\mathrm{GeV}}
\newcommand{\avg}[1]{\left\langle #1\right\rangle}

\begin{document}
\title{Energy-independent neutrino dephasing from an ultralight axion-like background}
\author{B. A. Couto e Silva}
\email{brunoaces@ufmg.br}
\author{B. L. S\'anchez-Vega}
\email{bruce@fisica.ufmg.br}
\affiliation{Departamento de F\'isica, UFMG, Belo Horizonte, MG 31270-901, Brazil}

\begin{abstract}
We study neutrino oscillations in an ultralight axion-like particle (ALP)
dark-matter background with low-energy derivative couplings $g_i$ diagonal
in the neutrino mass basis. In vacuum, these couplings generate an
additional relative phase set by the field difference between production
and detection, which can vary from event to event as the background
oscillates. Averaging over these unresolved variations can then reduce the
oscillation contrast. When the field changes little during each flight and
the spread of induced phases is small, the leading suppression is
quadratic in distance and independent of neutrino energy. Matching
published reactor damping limits at representative baselines gives scales
near $0.83$ and $0.5\GeV^{-1}$ for the coupling differences in JUNO and
KamLAND, respectively, with $g_2=g_3$ and a local ALP density of
$\rho_a=0.4\GeV\,\mathrm{cm}^{-3}$. These scales vary as $\rho_a^{-1/2}$ and
remain nearly mass independent, for instance over $10^{-20}\lesssim
m_a/\mathrm{eV}\lesssim10^{-13}$ when the data sample the assumed
distribution of the ALP field. We compare two electroweak operators: a dimension-five current
involving both neutrinos and charged leptons, constrained by
charged-lepton flavor violation, and a dimension-seven operator that
selects neutrinos through Higgs fields and allows invisible Higgs decays.
For the latter, the Higgs bound on the coupling differences is about two
orders of magnitude below the JUNO scale, assuming a single operator valid
at the Higgs mass and Standard Model production. At lower energies, we
calculate pion and kaon ALP-emission rates and missing-mass spectra and compare them with PIENU and NA62 searches.
\end{abstract}

\maketitle

\section{Introduction}
\label{sec:intro}

Neutrino oscillations are an interferometric measurement of relative phases.
The energy splittings involved are tiny, but their effects become observable
because they are accumulated over macroscopic distances. Neutrino propagation
is therefore sensitive to weak backgrounds that distinguish the mass
eigenstates even when the associated local energy shift is negligible compared
with the neutrino energy.

Ultralight bosonic dark matter provides a natural realization of such a
background. Its large occupation number permits a local classical-field
description. The Galactic velocity distribution sets the time and distance
over which the field remains coherent
\cite{Arias2012,Hui2021,Centers2021}. The effect on neutrino oscillations
depends on the interaction: an ultralight field may change neutrino masses or
act as an additional propagation potential
\cite{Berlin2016,BrdarEtAl2018,DevMachadoMartinez2021}.

One such possibility is a derivative coupling between neutrinos and an
axion-like particle (ALP) field. To isolate its effect on the propagation
phases, we consider the case in which this coupling is diagonal in the
neutrino mass basis. The background can then change the relative phases of
the propagating mass eigenstates without converting one into another. As the
ALP field oscillates, neutrinos produced at different times may acquire
different additional phases.

Each neutrino then propagates through a particular realization of the field,
while an experiment builds its oscillation spectrum from many events. If the
ALP field changes during data taking and its phase is not measured, the
observed spectrum combines events with different phase shifts. The central
question is how this combination affects the measured interference pattern,
and under what conditions the effect can be distinguished from ordinary
neutrino oscillations.

Related questions have been explored in studies of time-dependent neutrino
masses and propagation potentials from ultralight fields
\cite{HuangNath2018,DevMachadoMartinez2021,LosadaEtAl2022,
Lichkunov2025}. Loss-of-interference effects have also been examined in
open-system treatments of neutrinos \cite{BenattiFloreanini2000}.
Reactor-neutrino measurements, in turn, provide quantitative tests of departures
from standard oscillation interference \cite{BeccariaTernes2026}.

We address these questions for the diagonal derivative coupling described
above, working in an effective theory below the electroweak scale. We first
follow the propagation of a neutrino between its production and detection
and identify the phase entering the oscillation amplitude. We then examine
what changes when an experiment cannot resolve the time dependence of the
ALP background. This lets us identify the interference terms that carry
information about the coupling and ask which measurements can resolve them.
Reactor-neutrino energy spectra provide a concrete setting in which to make
this connection and compare the predicted changes with data. From this
starting point, we consider how the prospects change at longer baselines,
where the propagation conditions may also be different.

Oscillation measurements probe the neutrino coupling at low energies. The
low-energy interaction does not, however, specify its electroweak origin.
We therefore also examine gauge-invariant embeddings that relate the
neutrino coupling to other processes. In a specified embedding, these
relations allow measurements of other processes to constrain the same
coefficients. The comparison then shows how the electroweak assumptions
restrict the low-energy parameter space.

The remainder is organized as follows. Section~\ref{sec:phase}
introduces the low-energy interaction and derives the phase acquired during
neutrino propagation. Section~\ref{sec:average} examines how unresolved
variations of the ALP background affect oscillation interference.
Section~\ref{sec:neutrinos} discusses neutrino measurements, from reactor
data to longer terrestrial and astrophysical baselines.
Section~\ref{sec:ew} examines electroweak embeddings and their
complementary signals. We conclude in Sec.~\ref{sec:conclusions}.
Appendix~\ref{app:propagation} gives the current identities and mass
corrections underlying the phase calculation, together with the finite-time
relation between field representations. Appendix~\ref{app:exposure}
derives the interference factor for a finite exposure with general
event-time weights. Appendix~\ref{app:decays} presents the Higgs-decay
calculation and the meson emission rates and missing-mass spectra.

\section{Low-energy interaction and propagation phase}
\label{sec:phase}

\subsection{The interaction and its flavor structure}
\label{sec:left}

We work in a low-energy effective field theory (LEFT) below the electroweak
scale and do not assume a specific ultraviolet completion. The interaction
relevant for neutrino propagation is
\begin{equation}
 \mathcal L_{a\nu}=(\partial_\mu a)\bar\nu_i\gamma^\mu P_Lg_{ij}\nu_j,
 \qquad g=g^\dagger .
 \label{eq:left}
\end{equation}
The rest of the theory contains the neutrino kinetic and mass terms, the free
ALP Lagrangian, and the usual weak interactions that produce and detect the
neutrinos. Here $a$ is a real pseudoscalar of mass $m_a$, the indices label
neutrino mass eigenstates, and $P_L=(1-\gamma^5)/2$. We use $\hbar=c=1$ and
metric $(+---)$. The coefficients $g_{ij}$ have dimension $-1$.

We adopt the overall sign in Eq.~\eqref{eq:left} as a convention. Reversing
that sign is equivalent to $g_{ij}\to-g_{ij}$ if the convention is changed
consistently throughout.

The derivative form preserves a constant shift of $a$; the ALP mass is a
soft breaking of that symmetry. For real $a$, hermiticity of the interaction
requires $g=g^\dagger$. We focus on real diagonal couplings,
$g_{ij}=g_i\delta_{ij}$.
This is a phenomenological flavor assumption, not a consequence of the
effective theory.
In vacuum it makes the background diagonal in the same basis as the mass
term. Each component can acquire a different phase, but the ALP does not
transfer population between mass eigenstates. The discussion of other field
representations below briefly allows off-diagonal couplings to establish
contact with the literature; the oscillation benchmark remains diagonal.

A contribution proportional to the identity gives the same leading phase to
all mass eigenstates and therefore drops out of oscillation probabilities.
We therefore use $\Delta g_{ij}=g_i-g_j$ as the oscillation parameters.
There are only two independent differences. Keeping them separate from a
common coefficient will also be useful when comparing oscillations with
inclusive decay rates, which can depend on all the $g_i$.

Equation~\eqref{eq:left} specifies a neutrino-only ALP interaction in the
low-energy theory; it does not fix a coefficient for a charged-lepton or
Higgs operator. Those relations will be supplied by the embeddings in
Sec.~\ref{sec:ew}. The usual requirements on an effective theory still apply:
it must be used below its domain of validity, and any process implied by its
specified interactions remains a physical test. In particular, low-energy
meson emission need not wait for an electroweak completion.

\subsection{From the energy shift to the relative phase}
\label{sec:energy}

The derivative interaction changes the phases accumulated by the neutrino
mass eigenstates. We derive that phase directly from the
low-energy theory of Eq.~\eqref{eq:left}. We retain both temporal and spatial
derivatives of the ALP field and take the ultrarelativistic limit after
calculating the energy shift.

For a diagonal coupling, the equation of motion of the mass eigenstate
$\nu_i$ is
\begin{equation}
\left[
\mathrm{i}\gamma^\mu\partial_\mu
-m_i
+g_i(\partial_\mu a)\gamma^\mu P_L
\right]\nu_i=0.
\label{eq:dirac}
\end{equation}

Multiplication by $\gamma^0$ identifies the background perturbation as
$-g_i[(\partial_ta)P_L+(\bm\alpha\cdot\nabla a)P_L]$, where
$\alpha^k=\gamma^0\gamma^k$. The weak interaction produces predominantly
the negative-helicity neutrino. Its left-handed probability is $(1+v_i)/2$,
with $v_i=|\mathbf p|/E_i$, and its longitudinal current points along
$\widehat{\mathbf n}=\mathbf p/|\mathbf p|$. The first-order energy shift is
\begin{equation}
 \delta E_i=-g_i\frac{1+v_i}{2}
 (\partial_ta+\widehat{\mathbf n}\cdot\nabla a)
 \simeq -g_i\frac{\dd a}{\dd t}.
 \label{eq:shift}
\end{equation}
The last expression uses $v_i\simeq1$, so that the derivative is taken along
the neutrino trajectory. Appendix~\ref{app:propagation} gives the spinor
projection and the corrections for Dirac and Majorana masses. Their leading
result agrees for the same normalized left-handed current. Equation~\eqref{eq:shift}
is also the low-energy counterpart of the axial propagation potential in
Ref.~\cite{HuangNath2018}.

This form makes the accumulation of phase simple. Let $P=(t_P,\mathbf x_P)$
and $D=(t_D,\mathbf x_D)$ be the production and detection events. Integrating
the shift between them gives
\begin{equation}
 \begin{aligned}
 \Phi_i^{(a)}&=\int_{t_P}^{t_D}\delta E_i\dd t
             \simeq-g_i(a_D-a_P),\\
 \Delta\Phi_{ij}^{(a)}&=-\Delta g_{ij}(a_D-a_P).
 \end{aligned}
 \label{eq:phase}
\end{equation}
Here and below the equality for the relative ALP phase refers to the leading
ultrarelativistic order. It is the change of the field between the two events
that matters; a constant field gives no extra phase. Unlike a change of mass,
this contribution does not come with the kinematic factor $1/E$.

A universal shift \(g_i\to g_i+g_0\) adds the same endpoint phase to all
components and therefore multiplies the flavor amplitude by an overall phase.
At leading ultrarelativistic order, oscillation probabilities depend only on
the differences \(\Delta g_{ij}\).

The interfering mass components are evaluated at the same production and
detection events, as in the standard oscillation amplitude. We treat these
processes as localized and assume that the neutrino wave packets remain
sufficiently overlapped during propagation to follow a common semiclassical
trajectory. A loss of overlap due to their different group velocities would
be a separate source of decoherence.

The local expansion assumes $|g_i\partial a|\ll E_i$ and
$\omega_a,|\mathbf k_a|\ll E_i$, where $\omega_a$ and $\mathbf k_a$ are
background frequencies and momenta. These conditions keep the background
weak and slowly varying on microscopic propagation scales. They do not
require the integrated ALP phase to be perturbatively small, since a small
instantaneous energy shift may accumulate over a macroscopic baseline.
A universal component of $g_i$, which is invisible
to leading oscillation probabilities, must satisfy these conditions
independently.

Over the full flight, the accumulated corrections to the relative phase must
also remain small. Writing $L$ for the source--detector distance and $E$
for the common neutrino energy at leading ultrarelativistic order, we
retain the ordinary vacuum phase $m_i^2L/(2E)$ and require the omitted
relative background phase to be negligible; an explicit remainder is
given in Appendix~\ref{app:propagation}. Within this accuracy,
Eq.~\eqref{eq:phase} retains the change of the field over the full flight.

\subsection{The same phase after a field redefinition}
\label{sec:ibp}

Since the interaction is derivative, it can be integrated by parts. For
$J_i^\mu=\bar\nu_i\gamma^\mu P_L\nu_i$, its contribution to the action is
equivalent to $-g_i a\partial_\mu J_i^\mu$, up to a total derivative under
the usual boundary conditions. A coupling to an exactly conserved current
would then be redundant. The current of an individual neutrino field is not,
however, conserved by the complete theory used in an oscillation experiment:
its mass term and its weak production and detection interactions must also
be included. Unitary propagation with no transitions does not establish
that separate current conservation.

This can be seen without calculating a new propagation potential. Make the
local change of field
\begin{equation}
 \nu_{iL}(x)=e^{\ii g_i a(x)}\chi_{iL}(x).
 \label{eq:rotation}
\end{equation}
The derivative acting on the exponential in the kinetic term cancels the
explicit ALP current. The weak charged current still contains the neutrino,
so its mixing coefficient becomes $U_{\alpha i}e^{\ii g_i a(x)}$, where
$U$ is the neutrino mixing matrix. For production in flavor $\alpha$ and
detection in flavor $\beta$, with $\alpha,\beta\in\{e,\mu,\tau\}$,
mass component $i$ carries the conjugate factor at $P$ and the factor at $D$:
\begin{equation}
 \begin{aligned}
 &(U_{\alpha i}^*e^{-\ii g_i a_P})
 (U_{\beta i}e^{\ii g_i a_D})\\
 &\hspace{1cm}=U_{\alpha i}^*U_{\beta i}
 e^{\ii g_i(a_D-a_P)}.
 \end{aligned}
 \label{eq:endpointvertices}
\end{equation}
This is precisely the leading phase obtained from Eq.~\eqref{eq:shift}.
Removing it from propagation has moved it to the vertices, not removed it
from the measurement.

For Dirac neutrinos, the same change of fields makes the mass term
$-m_i\bar\chi_i e^{-\ii g_i a\gamma^5}\chi_i$. Its effect completes the
subleading mass corrections; it does not cancel the leading vertex factor.
The analogous two-component Majorana transformation and the full current
divergence are given in Appendix~\ref{app:propagation}. Using only the free
neutrino equations of motion in the integration by parts would omit the
weak-interaction terms responsible for Eq.~\eqref{eq:endpointvertices}.
This is the usual requirement to transform all operators consistently when
using field redefinitions in an effective theory~\cite{CriadoPerezVictoria2019}.

A neutrino produced in flavor $\alpha$ is the coherent state
$|\nu_\alpha\rangle=\sum_i U_{\alpha i}^*|\nu_i\rangle$. Including the
vacuum phase, the amplitude in either representation is
\begin{equation}
 \mathcal A_{\alpha\beta}[a]=
 \sum_i U_{\alpha i}^*U_{\beta i}
 e^{-\ii m_i^2L/(2E)}e^{\ii g_i(a_D-a_P)}.
 \label{eq:amplitude}
\end{equation}
The production and detection events are common to the interfering
components. Pulling out their local field phases assumes that $g_i a$
changes negligibly across each coherent vertex region. For an extended
source those phases stay inside the source integrals.
The ALP phase is included once: through the derivative-basis propagator or
through the transformed propagation and vertices, not through both.

\subsection{Connection to fluctuating-field treatments}
\label{sec:redfield}

For the diagonal coupling considered above, the ALP changes the relative
phases of the neutrino mass components without transferring population
between them. Off-diagonal couplings can also induce transitions from one
mass component to another. To relate these effects to fluctuating-field
treatments, we extend the same field redefinition to a general Hermitian
matrix $g$ and follow how the mass splittings enter the transformed
interaction.

Let $\psi_\nu$ collect the amplitudes of the neutrino mass components. In
vacuum, at leading ultrarelativistic order, they obey
\[
 \ii\dot\psi_\nu=(H_0-g\dot a)\psi_\nu,
 \qquad H_0=\frac{M^2}{2E}.
\]
Here $M^2=\diag(m_1^2,m_2^2,m_3^2)$, $E$ is the neutrino energy, and
the dot denotes a derivative along the trajectory.

The derivative potential can be absorbed into the matrix version of
Eq.~\eqref{eq:rotation}: $\psi_\nu=R\psi_\chi$, with
$R=e^{\ii ag}$. The amplitudes $\psi_\chi$ describe the same state in the
transformed representation. Substitution into the equation of motion adds
$-\ii R^\dagger\dot R=g\dot a$ to the Hamiltonian, canceling the
explicit derivative potential. The transformed amplitudes therefore obey
$\ii\dot\psi_\chi=H_\chi\psi_\chi$, with
\begin{equation}
 H_\chi=R^\dagger H_0R.
 \label{eq:representation}
\end{equation}
The background now enters through a rotation of the vacuum Hamiltonian.
If $g$ is diagonal in the mass basis, this rotation leaves $H_0$
unchanged. If $g$ connects states with different masses, it also produces
off-diagonal terms in $H_\chi$.

To identify the term linear in the field, expand
$R=1+\ii ag+\cdots$ and $R^\dagger=1-\ii ag+\cdots$.
Multiplying out $R^\dagger H_0R$ gives
$H_\chi=H_0+\ii a(H_0g-gH_0)+\cdots$.
Writing this as $H_\chi=H_0+a\mathcal J+\cdots$, we obtain
\begin{equation}
 \mathcal J=\ii[H_0,g],\qquad
 \mathcal J_{ij}=\ii\frac{\Delta m_{ij}^2}{2E}g_{ij}.
 \label{eq:J}
\end{equation}
Here $\mathcal J$ is the coefficient matrix of the linear interaction and
$\Delta m_{ij}^2=m_i^2-m_j^2$. Because $H_0$ is diagonal, its commutator
with $g$ multiplies each $g_{ij}$ by the vacuum energy difference between
the two states it connects. This is the origin of the factor $1/E$.
The higher-order terms needed when expanding probabilities are discussed
in Appendix~\ref{app:finite}.

Equation~\eqref{eq:J} reproduces the relativistic matrix element in
Eq.~(34) of Ref.~\cite{Lichkunov2025} for a left-handed current.\footnote{In
that notation, $C_V$ and $C_A$ are the vector and axial coupling matrices,
and $F$ is the scale used to normalize the derivative interaction. Our
left-handed current corresponds to $C_V/F=g/2$ and $C_A/F=-g/2$.}
That treatment averages the density matrices obtained from unitary
evolution in different realizations of the field. The fluctuation term
in the resulting evolution equation is quadratic in $\mathcal J$, so
the two factors of the vacuum energy splitting supply an explicit
$E^{-2}$ dependence.

The transition probability also depends on how the background varies
during a flight. Contributions accumulated at different times interfere,
so averaging the squared amplitude involves correlations of the background
at those times. For a stationary field, these depend only on the time
separation. When the relevant correlations decay on a time scale much
shorter than the flight and the transition probability remains small,
we can identify a contribution proportional to the flight time. Its
coefficient defines a transition rate. This rate depends on the field
power spectrum---the strength of fluctuations at each frequency---evaluated
at the vacuum energy splitting $\Delta m_{ij}^2/(2E)$. Changing the
neutrino energy changes this frequency, so the spectrum can add energy
dependence beyond the explicit $E^{-2}$ factor.
Appendix~\ref{app:finite} derives this limit for a differentiable field
and obtains the same rate in the derivative and rotated representations.

For the diagonal coupling studied here, $g$ commutes with $H_0$, so
$\mathcal J=0$ and $H_\chi=H_0$. The weak production and detection
vertices still carry the field phases. Their product contributes
$e^{\ii g_i(a_D-a_P)}$ to each mass component, as in
Eq.~\eqref{eq:amplitude}. These phases follow from the unexpanded rotation
and can be retained for arbitrary $g_i a$ within the ultrarelativistic
description. Thus the background can change interference without inducing
transitions between mass eigenstates. Combining production, propagation,
and detection for each field realization gives the same complete
probability in either representation.

Neutrinos produced at different times can encounter different field
values even when they follow the same path. The experimental average
then combines probabilities for flights with different additional
phases. The event times determine how these phases are sampled during
the observation, while each neutrino evolves over its own flight time.
This sampling can occur even when each flight is short compared with
the field correlation time. The phase average and the long-time
transition rate therefore need not have the same energy or distance
dependence.

\section{Unresolved field variations and oscillation contrast}
\label{sec:average}
\subsection{A coherent ALP mode}
\label{sec:mode}

An ultralight bosonic field with a large occupation number can be treated
locally as a classical background
\cite{Arias2012,GrahamRajendran2013,Hui2021}. The ALP field in a Galactic
halo is a superposition of modes with slightly different velocities and
frequencies. We first consider a single coherent mode as a simple
idealization,
\begin{equation}
 a(t,\mathbf x)=a_0\cos\xi(t,\mathbf x),\qquad
 \xi=\omega_at-\mathbf k_a\cdot\mathbf x+\theta .
 \label{eq:mode}
\end{equation}
Here $a_0$ is the amplitude and $\theta$ is a constant phase.
For a free mode, $\omega_a^2=m_a^2+|\mathbf k_a|^2$. Galactic halo modes
are nonrelativistic, with $v_a=|\mathbf k_a|/\omega_a$ of order $10^{-3}$,
so $\omega_a\simeq m_a$~\cite{GrahamRajendran2013,Gramolin2022}.

We allow the ALP to make up all or part of the local dark matter and use
$\rho_{\rm DM}=0.4\GeV\,\mathrm{cm}^{-3}$ as a reference density~\cite{deSalasWidmark2021}.
In this single-mode idealization, we assign the local ALP density $\rho_a$
to the mode. Its time-averaged energy density is $\rho_a=\omega_a^2a_0^2/2$,
which gives
\begin{equation}
 a_0=\frac{\sqrt{2\rho_a}}{\omega_a},\qquad
 \partial_ta=-\sqrt{2\rho_a}\sin\xi .
 \label{eq:normalization}
\end{equation}
The spatial derivative is $\nabla a=(\mathbf k_a/\omega_a)
\sqrt{2\rho_a}\sin\xi$ and is smaller by the halo velocity.

To evaluate the field difference in Eq.~\eqref{eq:phase}, we follow the
ultrarelativistic trajectory
$\mathbf x(t)=\mathbf x_P+\widehat{\mathbf n}(t-t_P)$.
The temporal and spatial dependence of the field then combine into the
frequency $\Omega=\omega_a-\mathbf k_a\cdot\widehat{\mathbf n}$.
At the order retained in the phase calculation, the flight time is $L$.
The field phase therefore changes from $\xi_P=\xi(t_P,\mathbf x_P)$ at
production to $\xi_P+\Omega L$ at detection. Taking the difference of the
two cosines gives
\begin{equation}
 a_D-a_P=-2a_0\sin\frac{\Omega L}{2}
                  \sin\left(\xi_P+\frac{\Omega L}{2}\right).
 \label{eq:fielddifference}
\end{equation}
Thus events emitted at different ALP phases acquire different additional
oscillation phases, even for the same neutrino energy and baseline.
Substituting this field difference into Eq.~\eqref{eq:phase} gives the
amplitude of the relative phase,
\begin{equation}
 A_{ij}=2|\Delta g_{ij}|\frac{\sqrt{2\rho_a}}{\omega_a}
             \left|\sin\frac{\Omega L}{2}\right|.
 \label{eq:phaseamplitude}
\end{equation}
This expression retains the change of the field during the flight. At fixed
$\rho_a$, the amplitude of $\partial_ta$ is independent of $m_a$. The
accumulated phase, however, still depends on the mass through
$1/\omega_a$ and the phase advance $\Omega L$.
The phase amplitude grows linearly with distance only for $|\Omega L|\ll1$.
For longer flights, the oscillating field can partly undo a phase accumulated
earlier; the endpoints, rather than an unrestricted power of $L$, determine
the answer. We next examine how events sampling different initial phases
contribute to the measured interference pattern.

\subsection{Averaging the measured interference}
\label{sec:coherentaverage}

For each event, propagation through the ALP background is unitary.
Events following the same path can nevertheless be produced at different
field phases $\xi_P$ and acquire different relative phases. If the
measurement does not resolve these variations, the observed probability
is obtained by averaging $|\mathcal A_{\alpha\beta}|^2$ over the phases
sampled by the events.

For the single mode considered above, assume that $\xi_P$ is sampled
uniformly. In the interference factor, let $\varphi$ denote this phase
with the fixed trajectory offset $\Omega L/2$ included and any overall
sign absorbed into a constant shift. The distribution of $\varphi$ is
also uniform, so
\begin{equation}
 \mathcal C_{ij}=\frac{1}{2\pi}\int_0^{2\pi}
 e^{\ii A_{ij}\sin\varphi}\dd\varphi=J_0(A_{ij}),
 \label{eq:bessel}
\end{equation}
where $A_{ij}$ is the phase amplitude in Eq.~\eqref{eq:phaseamplitude}
and $J_0$ is the Bessel function of the first kind of order zero. Its
integral representation gives the last equality~\cite{Watson1944}.
Substituting this average into the interference terms gives the vacuum
probability
\begin{equation}
 \begin{aligned}
 \overline P_{\alpha\beta}={}&
 \sum_i|U_{\alpha i}|^2|U_{\beta i}|^2\\
 &+2\Re\sum_{i>j}U_{\alpha i}^*U_{\beta i}
 U_{\alpha j}U_{\beta j}^*\mathcal C_{ij}
 e^{-\ii\Delta m_{ij}^2L/(2E)} .
 \end{aligned}
 \label{eq:probability}
\end{equation}
Here $\Delta m_{ij}^2=m_i^2-m_j^2$, and the bar denotes the average
over the sampled ALP phases. The same expression applies to other field
statistics, with the corresponding interference factor.

The first sum contains the separate contributions of the mass states
and is unchanged by the field average. Each term in the second sum
describes interference between two components and is multiplied by
$J_0(A_{ij})$. Contributions with different relative phases can partially
cancel when the events are combined, changing the observed oscillation
contrast~\cite{Schlosshauer2007}. The probability in each mass state
and the total probability remain unchanged.

For antineutrinos, the ALP-induced phase has the opposite sign. Since
$J_0(-A)=J_0(A)$, neutrinos and antineutrinos following the same path
and sampling the same uniform phase distribution have the same ALP
interference factors in vacuum. The antineutrino probability is obtained
from Eq.~\eqref{eq:probability} with $U\to U^*$. Measurements that
resolve the time variation can retain sensitivity to the sign of the
additional phase.

\subsection{Leading behavior of the interference factor}
\label{sec:quadratic}

The interference factor takes a simple form when the ALP field varies little
during a neutrino flight. For $|\Omega L|\ll1$, the field change is
approximately its derivative along the trajectory at production multiplied
by $L$. Since the amplitude of that derivative is fixed by $\rho_a$,
Eq.~\eqref{eq:phaseamplitude} becomes
$A_{ij}\simeq|\Delta g_{ij}|\sqrt{2\rho_a}L$, neglecting the
halo-velocity correction.

A further simplification is possible when the induced phase spread is also
small, $A_{ij}\ll1$. This is a separate condition: the phase amplitude
depends on the coupling and density as well as the field variation during
the flight. For the uniform phase sampling in Eq.~\eqref{eq:bessel}, the
additional relative phase has zero mean. Since $\avg{\sin^2\varphi}=1/2$,
its mean square is $\avg{(\Delta\Phi_{ij}^{(a)})^2}=A_{ij}^2/2$.
Because the mean is zero, this mean square is also the variance, which
measures the spread of the additional phases among events.

The expansion $J_0(A_{ij})=1-A_{ij}^2/4+\mathcal O(A_{ij}^4)$ shows
that the leading reduction of interference is one half of this variance.
At this order, the interference factor can be written as
\begin{equation}
 \mathcal C_{ij}\simeq
 \exp\!\left[-\frac{(\Delta g_{ij})^2\rho_aL^2}{2}\right].
 \label{eq:L2}
\end{equation}
This exponential reproduces the Bessel factor through quadratic order;
the two expressions differ at order $A_{ij}^4$.

Since $A_{ij}$ grows linearly with $L$ in this regime, the phase variance
and the leading reduction of contrast grow as $L^2$.
The $E^0$ dependence follows from the derivative potential, which has no
leading factor of $1/E$. At fixed density, the factor $1/\omega_a$ in the
field amplitude also cancels the factor $\Omega\simeq\omega_a$ from the
small-$|\Omega L|$ expansion of the sine. This accounts for the absence
of $m_a$ in the approximate result. The standard energy-dependent vacuum
phase in Eq.~\eqref{eq:probability} remains unchanged.

A scalar interaction that changes the mass matrix instead changes the
relativistic phase through $M^2/(2E)$. Squaring that phase spread can then
produce $L^2/E^2$~\cite{Berlin2016,DevMachadoMartinez2021,Airoldi2026}.
The different energy behavior reflects how the field enters propagation,
not the act of averaging itself.

For large $A_{ij}$, $J_0$ can cross zero and change sign. The interference
term can then reverse sign, and a positive exponential no longer describes
the result. When $|\Omega L|$ is not small, the sine in
Eq.~\eqref{eq:phaseamplitude} must be retained even if the phase amplitude
is small. The distance dependence then need not be quadratic. The amplitude
is bounded by $2|\Delta g_{ij}|\sqrt{2\rho_a}/\omega_a$ for the specified mode.

\subsection{From a mode to a halo and a finite data set}
\label{sec:halo}

The Bessel factor is exact for a sinusoid with fixed amplitude and uniformly
sampled phase. A Galactic halo instead contains modes with different
velocities, whose superposition gives a local field with fluctuating
amplitude and phase~\cite{Hui2021,Centers2021,Gramolin2022}. The phase
acquired by each neutrino still depends on $a_D-a_P$. What changes is the
distribution of this field difference, and hence the average over the
additional phases.

For fixed production and detection events, averaging over the field
realizations described by a chosen model gives
\begin{equation}
 \mathcal C_{ij}=\avg{e^{\ii\Delta g_{ij}(a_D-a_P)}}.
 \label{eq:C}
\end{equation}
Here the brackets denote a theoretical average over those realizations.
This is the characteristic function of the field difference: it determines
how the distribution of additional phases modifies the interference. Whether
a finite set of events samples this distribution is a separate question,
to which we return below.

Suppose the field is Gaussian with zero mean. Then $a_D-a_P$ is also
Gaussian, with zero mean and variance $\avg{(a_D-a_P)^2}$. Averaging the
phase exponential with this Gaussian distribution yields
\begin{equation}
 \mathcal C_{ij}=\exp\!\left[-\frac{(\Delta g_{ij})^2}{2} \avg{(a_D-a_P)^2}\right].
 \label{eq:gaussian}
\end{equation}
The exponent is minus one half of the variance of the additional relative
phase. A narrow phase distribution leaves the interference nearly unchanged;
a broader one produces a stronger cancellation in the average. This result
is exact for the assumed Gaussian statistics and does not require a small
phase spread.

To determine that spread, we need both the field variance at each end of
the path and the relation between the two values. Expanding the square gives
\begin{equation}
 \begin{aligned}
 \avg{(a_D-a_P)^2}={}\avg{a_D^2}+\avg{a_P^2}-2\avg{a_Da_P}.
 \end{aligned}
 \label{eq:covariance}
\end{equation}
The first two terms give the field variances at detection and production,
while the last measures the correlation between the two values. Even in a
Gaussian field, those values need not be independent.

When the field changes little during a flight,
$a_D-a_P\simeq L(\dd a/\dd t)_P$, where the derivative follows the
neutrino trajectory. The variance is consequently
$\avg{(a_D-a_P)^2}\simeq L^2\avg{(\dd a/\dd t)^2}\simeq\rho_aL^2$
at leading order in the halo velocity. Substitution into
Eq.~\eqref{eq:gaussian} recovers Eq.~\eqref{eq:L2}. The Gaussian and
fixed-amplitude descriptions therefore give the same leading reduction
when the induced phase spread is small. Beyond that order their different
distributions matter: the Gaussian factor remains positive and has none of
the zeros or sign changes of $J_0$.

For fixed, finite field variances at production and detection,
Eq.~\eqref{eq:covariance} also bounds the variance of their difference.
If the two field values become uncorrelated, this variance approaches the
sum of the two individual variances, or $2\avg{a^2}$ when they are equal,
rather than continuing to grow as $L^2$. A perfectly coherent mode can
instead retain an oscillatory dependence on distance. The separate
variances must be kept when source and detector occupy regions with
different field densities. Spatial separation can also reduce the
correlation: for characteristic halo speeds $v_a$, the spatial coherence
scale is of order $(m_av_a)^{-1}$. This compares field values at the same
time; the production--detection correlation also includes the evolution
during the flight~\cite{GrahamRajendran2013,Centers2021,Gramolin2022}.

The connection to data requires distinguishing one neutrino flight from the
full observation. Let $T$ be the elapsed span of data taking. In addition
to the ALP period, approximately $2\pi/m_a$, the local field has a coherence
time $\tau_{\rm coh}\sim(m_av_a^2)^{-1}$~\cite{GrahamRajendran2013,Centers2021,Gramolin2022}.
For times much shorter than $\tau_{\rm coh}$, the local field is
approximately sinusoidal, with nearly fixed amplitude and phase offset.
It continues to oscillate on the ALP period; the coherence time describes
the slower changes caused by the small frequency differences between modes.

For events along a fixed path within a spatially coherent region, a nearly
constant additional phase shifts the oscillation pattern without reducing
its contrast. If instead the events cover many ALP periods while the
amplitude remains nearly fixed, uniform sampling of the phase gives
Eq.~\eqref{eq:bessel}. Thus $m_aL\ll1$ controls the field variation during
one flight, whereas $m_aT\gg1$ allows many oscillations of the field during
data taking. Both can hold simultaneously. The event-time weights determine
whether those oscillations are sampled uniformly.

Even when $T\ll\tau_{\rm coh}$, the local amplitude is a random property
of the halo realization. Observing many cycles samples its phase, not the
distribution of possible amplitudes. The Bessel result is therefore
conditional on the amplitude present during the run. A statistical analysis
may treat an unknown amplitude as a parameter with the distribution assigned
by the halo model~\cite{Centers2021}. If $T\gg\tau_{\rm coh}$, events
can sample different local amplitudes and phases. If the halo statistics
remain stable and the sampling is representative, their average can approach
Eq.~\eqref{eq:gaussian} for the Gaussian field model. Around
$T\sim\tau_{\rm coh}$, the prediction must retain how the field values
are correlated between events and the
relative weight of those events.

Appendix~\ref{app:exposure} gives the average with an explicit event-time
weight. A finite average can change the oscillation phase as well as its
contrast. Assigning a distribution to an unknown initial ALP phase describes
uncertainty about the realization; averaging events describes the phases
actually sampled. The reactor comparisons below state the conditions under
which the simple field averages are used.

\section{Neutrino probes of the low-energy interaction}
\label{sec:neutrinos}

To connect the ALP-induced phase with neutrino measurements, we need to
identify which interference terms contribute to the detected flavor and
how well the experiment can resolve their oscillation pattern. Reactor
spectra provide a concrete setting for this comparison. We then examine
how the background density and the propagation conditions affect the
interpretation of the signal over longer paths.

\subsection{Flavor dependence of the oscillation signal}
\label{sec:flavor}

Reactor experiments measure the survival probability of electron
antineutrinos. Because an electron antineutrino is a superposition of the
three mass eigenstates, this probability depends on their relative phases
and on the mixing coefficients $U_{ei}$. For a symmetric distribution of
the ALP-induced phase, the factors $\mathcal C_{ij}$ are real, and
Eq.~\eqref{eq:probability} gives the vacuum survival probability
\begin{equation}
 \begin{aligned}
 \overline P_{ee}={}&\sum_i|U_{ei}|^4\\
 &+2\sum_{i>j}|U_{ei}|^2|U_{ej}|^2\mathcal C_{ij}
 \cos\frac{\Delta m_{ij}^2L}{2E}.
 \end{aligned}
 \label{eq:reactorprob}
\end{equation}
The bar denotes the average over the ALP phases sampled by the events,
at fixed energy $E$ and propagation distance $L$.
The first sum contains the separate contributions of the mass states and
is unchanged by the ALP phase. Each term in the second sum describes
interference between a pair of states, with weight
$2|U_{ei}|^2|U_{ej}|^2$ and field-average factor $\mathcal C_{ij}$.

These weights determine how strongly a change in each interference term
affects $\overline P_{ee}$. The squared magnitudes in the electron row
of $U$ are determined by the mixing angles $\theta_{12}$ and $\theta_{13}$:
$|U_{e1}|^2=\cos^2\theta_{12}\cos^2\theta_{13}$,
$|U_{e2}|^2=\sin^2\theta_{12}\cos^2\theta_{13}$, and
$|U_{e3}|^2=\sin^2\theta_{13}$. To illustrate their relative sizes, we use
$\sin^2\theta_{12}=0.307$ and $\sin^2\theta_{13}=0.022$,
representative rounded values from the NuFIT 6.0 analysis for normal
ordering~\cite{NuFIT2024}. The weights of the $21$, $31$, and $32$
terms are then approximately $0.407$, $0.0298$, and $0.0132$.
The $21$ interference has the largest weight; the other two contain the
small electron component of state $3$.

The three terms also oscillate on different scales. The smaller splitting
$\Delta m_{21}^2$, known as the solar mass splitting because it governs
solar-neutrino oscillations, gives the slow component of the reactor
spectrum. The larger splittings $|\Delta m_{31}^2|$ and
$|\Delta m_{32}^2|$ produce the two faster variations with $L/E$.
Measuring the contrast of the slow component therefore probes
$\Delta g_{21}$.

Which terms are changed by the ALP follows from the coupling differences.
There are only two independent differences, since
\begin{equation}
 \Delta g_{32}=\Delta g_{31}-\Delta g_{21}.
 \label{eq:flavor}
\end{equation}
When two mass eigenstates have equal couplings, they acquire the same
additional phase. Their interference is therefore unchanged by the ALP.
For $g_2=g_3$, the equal differences
$\Delta g_{21}=\Delta g_{31}$ modify the $21$ and $31$ terms, while
$\mathcal C_{32}=1$. For $g_1=g_3$, the affected pairs are $21$ and
$32$, while $\mathcal C_{31}=1$. Both choices act on the slow component.
The pattern $g_1=g_2$ instead leaves $\mathcal C_{21}=1$ and modifies
only the two fast terms.

The measured spectrum also averages over neutrino energies and reactor
distances. A fast oscillation can pass through several maxima and minima
within one reconstructed-energy interval or across the contributing
baselines, so its interference term can largely cancel in that average.
In the limit where both fast terms, including their ALP factors, average
to zero, $g_2=g_3$ and $g_1=g_3$ give the same averaged spectrum for
the same $|\Delta g_{21}|$ and field distribution. Only their common
effect on the slow component remains. The choice $g_1=g_2$ leaves that
component unchanged. Sensitivity to this pattern through vacuum dephasing
therefore requires some of the fast interference to remain observable.

Within the regime of Eq.~\eqref{eq:L2}, a longer baseline increases the
spread of ALP-induced phases. The mixing weights and the interference
retained in the spectrum determine how this changes the measured signal.

\subsection{Reactor measurements: JUNO and KamLAND}
\label{sec:reactors}

We now use published limits on oscillation damping to identify the
corresponding ALP coupling scales. These analyses fit the damping together
with the standard oscillation parameters. We use their energy-independent
parametrization, in which the interference between mass states $i$ and $j$
is multiplied by $e^{-\gamma_{ij}L}$
\cite{DeRomeri2023,BeccariaTernes2026}. The nonnegative fit coefficient
$\gamma_{ij}$ has units of inverse length, or energy in natural units.
The choice $\gamma_{21}=\gamma_{31}\equiv\gamma$, $\gamma_{32}=0$
modifies the same two terms as our coupling pattern $g_2=g_3$, for which
we write $|\Delta g|=|\Delta g_{21}|=|\Delta g_{31}|$. The exponent of the
reference factor is linear in $L$, whereas the leading ALP reduction of
contrast is quadratic in $L$ when the field changes little during a
flight. We compare them by finding the coupling that gives the same
interference factor at a representative distance, denoted by $L_*$.
We use the vacuum ALP factors for this conversion; Sec.~\ref{sec:matter}
describes how matter enters the propagation.

We adopt $L_*=52.5$ km for JUNO and $180$ km for KamLAND. For JUNO,
the eight principal reactor cores lie between 52.11 and 52.82 km from
the detector in the nominal configuration of
Ref.~\cite{ForeroParkeTernesFunchal2021}. For a geometric check, we take
the antineutrino output of each core to be proportional to its thermal
power. A more distant core contributes a smaller flux because its
neutrinos spread over an area that grows as the square of the distance.
We use these fluxes as weights when averaging the squared distances,
then take the square root. The result, $52.54$ km, supports $52.5$ km
as a reference for the main reactor contribution. For KamLAND, we adopt
the flux-weighted average of $180$ km used to present its oscillation
data~\cite{KamLAND2008}. Its reactors span a broader range of distances,
so one value represents the sample less closely
\cite{KamLAND2008,KamLAND2013}.

The events are detected through inverse beta decay,
$\bar\nu_e+p\to e^++n$, with the visible positron signal used to estimate
the incident antineutrino energy $E$~\cite{JUNOPhysics2022,KamLAND2013}.
Finite resolution allows different incident energies to enter the same
reconstructed-energy interval $b$. Contributions from reactors $r$ at
distances $L_r$ add to give the expected signal
\begin{equation}
 N_b=\sum_r\int\dd E\,K_{br}(E)\overline P_{ee}(E,L_r),
 \label{eq:reactorspectrum}
\end{equation}
where $K_{br}(E)$ includes the unoscillated flux, interaction cross
section, exposure, number of target protons, detection efficiency, and
probability of assigning an event of energy $E$ to interval $b$.
The field average uses the event-time weights for each contribution.
A spectral fit retains each $L_r$ in this sum; our conversion compares
only the interference factors at $L_*$.

Let $\gamma_{\max}$ be the published upper limit for the chosen pattern.
At $L_*$ it corresponds to the factor $e^{-\gamma_{\max}L_*}$. For a
Gaussian field with $m_aL_*\ll1$, the variance of $a_D-a_P$ in
Eq.~\eqref{eq:gaussian} is approximately $\rho_aL_*^2$, giving
$\exp[-(\Delta g)^2\rho_aL_*^2/2]$. Equating the exponents determines
the corresponding coupling. For the sinusoidal background of
Sec.~\ref{sec:mode}, with its amplitude held fixed and its phase
uniformly sampled, we instead find the phase amplitude $A_*$ that gives
the same factor through $J_0(A_*)$. The two conversions are
\begin{equation}
 \begin{aligned}
 |\Delta g|_{\rm G}&=\sqrt{\frac{2\gamma_{\max}}{\rho_aL_*}},\\
 J_0(A_*)&=e^{-\gamma_{\max}L_*},\qquad
 |\Delta g|_{\rm mode}=\frac{A_*}{\sqrt{2\rho_a}L_*}.
 \end{aligned}
 \label{eq:recast}
\end{equation}
The subscripts G and mode distinguish the two distributions of field
values. For the mode, we take the solution reached by increasing the
coupling from zero, before the first zero of $J_0$. Both conversions
require little field variation during a flight and sampling of the
distribution used in the average. The Gaussian average follows from its
variance, while the mode conversion retains the full Bessel function.

The published limits also depend on the information used to constrain
the ordinary oscillation parameters. In particular, $\theta_{12}$ sets
the weight of the slow $21$ interference term. Varying this angle can
partly compensate for damping in the reactor spectrum. The reference
analyses use solar-neutrino measurements to constrain $\theta_{12}$ and
help separate the two effects~\cite{DeRomeri2023,BeccariaTernes2026}.
We retain this solar input when converting their limits.

The updated JUNO sample contains 8294 inverse beta decay candidates
collected over 207.2 live days~\cite{JUNO207Days2026}. For the pattern
above, Ref.~\cite{BeccariaTernes2026} obtains
$\gamma<2.87\times10^{-22}\GeV$ at 90\% confidence with the solar
constraint. At $L_*=52.5$ km, this limit corresponds to each affected
interference term retaining about 92.6\% of its undamped coefficient.
Using $\rho_a=0.4\GeV\,\mathrm{cm}^{-3}$ in Eq.~\eqref{eq:recast},
we obtain $|\Delta g|_{\rm G}\simeq0.838\GeV^{-1}$ for the Gaussian
field and $|\Delta g|_{\rm mode}\simeq0.830\GeV^{-1}$ for the sinusoid.

For KamLAND we use the approximate limit
$\gamma_{\max}\simeq4\times10^{-22}\GeV$ summarized in
Refs.~\cite{DeRomeri2023,BeccariaTernes2026}, with its external solar
information and $L_*=180$ km. Although KamLAND allows a larger
$\gamma$, its longer path gives a smaller value of
$\gamma_{\max}/L_*$, the combination entering the Gaussian conversion.
Table~\ref{tab:reactors} collects the inputs and the resulting coupling
scales. The comparison thus uses both the propagation distance and the
contrast allowed by the measured spectrum.

\begin{table}[htbp]
\caption{Reference damping limits and the coupling scales obtained from
Eq.~\eqref{eq:recast}, for $g_2=g_3$ and
$\rho_a=0.4\GeV\,\mathrm{cm}^{-3}$. The input $\gamma_{\max}$ values
are the published 90\%-confidence limits with external solar
information, quoted in units of $10^{-22}$ GeV. The last two columns
are in GeV$^{-1}$ and use, respectively, a Gaussian field and a single
sinusoid of fixed amplitude with uniform phase sampling. The JUNO entry
uses the analysis of the 207.2-live-day sample. The conversions assume
$m_aL_*\ll1$ and effective sampling of the stated field distribution;
the confidence level refers to the reference damping fits.}
\label{tab:reactors}
\begin{ruledtabular}
\begin{tabular}{lrrrr}
 & $L_*$ [km] & $\gamma_{\max}$ & $|\Delta g|_{\rm G}$ & $|\Delta g|_{\rm mode}$\\
JUNO & 52.5 & 2.87 & 0.838 & 0.830\\
KamLAND & 180 & $\simeq4$ & 0.534 & 0.510
\end{tabular}
\end{ruledtabular}
\end{table}

For small phase spreads, the two field distributions give the same
leading reduction of contrast. Their full averages give coupling scales
that differ by about 1\% for JUNO and 5\% for KamLAND. The larger
allowed phase spread at KamLAND makes the difference between the two
averages more pronounced. These percentages quantify the dependence of
the conversion on the assumed field distribution.

The alternative pattern $g_1=g_3$ changes the $21$ and $32$ terms and
corresponds to $\gamma_{21}=\gamma_{32}$, $\gamma_{31}=0$ in the
reference parametrization. Like $g_2=g_3$, it modifies the dominant
slow component. Reference~\cite{BeccariaTernes2026} reports that it
would give similar results and does not analyze it separately. The same
study also constrains the pattern $\gamma_{31}=\gamma_{32}\equiv\gamma$,
$\gamma_{21}=0$, obtaining $\gamma<6.73\times10^{-21}\GeV$ at 90\%
confidence for the energy-independent case. This corresponds to
$g_1=g_2$ and modifies only the fast terms. The weaker limit reflects
how much of that fast interference is resolved by the sample, as
discussed in Sec.~\ref{sec:flavor}.

For $g_1=g_2$, the nonzero differences have magnitude
$|\Delta g_{31}|=|\Delta g_{32}|$. At the same JUNO distance and density
used in Table~\ref{tab:reactors}, the reference factor is
$e^{-\gamma_{\max}L_*}\simeq0.167$. Equation~\eqref{eq:recast} then gives
\[
 |\Delta g_{31}|_{\rm G}\simeq4.06\GeV^{-1},\qquad
 |\Delta g_{31}|_{\rm mode}\simeq3.18\GeV^{-1}.
\]
The sinusoidal value uses $A_*\simeq2.10$ on the first branch of $J_0$.
These scales are about 4.8 and 3.8 times the corresponding values for
$g_2=g_3$. Both have the same $\rho_a^{-1/2}$ dependence and sampling
conditions as the entries in Table~\ref{tab:reactors}. Here too, the
90\% confidence level refers to the reference fit for $\gamma$.

The nearly mass-independent coupling scales apply when the field changes
little between production and detection, $m_aL_*\ll1$, while the
observations sample the variations used in the average. These conditions
involve different times. The JUNO data were collected from 30 August
2025 to 18 May 2026, spanning about 261 calendar days
\cite{JUNO207Days2026}. Ten ALP periods within this span correspond to
$m_a\simeq1.8\times10^{-21}$ eV. This illustrates the available temporal
coverage; the actual average depends on the distribution of event times.
Many sampled periods can give the sinusoidal phase average when the
amplitude remains approximately constant. The Gaussian conversion also
requires the assumptions about the halo realization described in
Sec.~\ref{sec:halo}.

At larger masses the field changes more during a flight, and the sine
in Eq.~\eqref{eq:phaseamplitude} must be retained. For JUNO, doing so
changes the inferred single-mode coupling by less than 5\% relative to
its value in Table~\ref{tab:reactors} for
$m_a\lesssim4\times10^{-12}$ eV, provided the assumed phase sampling
holds. At this upper mass $m_aL_*\simeq1.06$, so the comparison measures
the accuracy of using Eq.~\eqref{eq:recast} beyond the strict
$m_aL_*\ll1$ limit. For KamLAND,
$L_*^{-1}\simeq1.1\times10^{-12}$ eV sets the mass at which the field
phase changes by about one radian during the representative flight;
Eq.~\eqref{eq:recast} requires masses well below this value.
Phase sampling must also hold over KamLAND's own data-taking interval.
The two conversions can be compared where both experiments meet their
flight and sampling conditions.

The solar information used in these conversions also needs to be
interpreted within the ALP model. The background contributes to
solar-neutrino propagation and may change the inferred $\theta_{12}$.
For illustration, consider the difference between the ALP potentials of
mass states 2 and 1. With the sinusoidal normalization,
$|\Delta g|=0.83\GeV^{-1}$ and the reference density give an amplitude
of about $2.1\times10^{-21}$ GeV. At $E=10$ MeV, this is about $0.55$
times the vacuum energy splitting $\Delta m_{21}^2/(2E)$, for
$\Delta m_{21}^2=7.5\times10^{-5}\,\mathrm{eV}^2$. The potential is
negligible compared with the neutrino energy, yet can be relevant to the
splittings that govern oscillations. Its effect on solar measurements
depends on the combined ALP and matter evolution through the Sun.
Our conversions retain the solar constraints of the reference fits.

The table identifies the coupling scales that reproduce the allowed
reference interference at the chosen distances and density. A fit of the
ALP model to the measured spectra would use Eq.~\eqref{eq:reactorspectrum}
with the individual reactor distances, field statistics, and event-time
weights, treating matter propagation and the solar information within
the same model. In the comparison made here, KamLAND's longer path
corresponds to a smaller coupling scale. Both measurements probe
$|\Delta g|\sqrt{\rho_a}$; we next examine that density dependence.

\subsection{Dependence on the local ALP density}
\label{sec:density}

The coupling scales in Table~\ref{tab:reactors} were obtained by assigning
the full reference local dark-matter density,
$\rho_{\rm DM}=0.4\GeV\,\mathrm{cm}^{-3}$, to the ALP. If ALPs make up
only part of that density, the field amplitude is smaller. A larger
coupling is then needed to produce the same change in oscillation contrast.

We write the local ALP fraction as
$F_{\rm loc}\equiv\rho_a/\rho_{\rm DM}$, so that
$\rho_a=F_{\rm loc}\rho_{\rm DM}$. At fixed ALP mass, baseline, and
velocity distribution, the spread of the additional phase is proportional
to $|\Delta g|\sqrt{\rho_a}$ for either field description used in
Eq.~\eqref{eq:recast}. Keeping this product fixed therefore preserves
the interference factor. Denoting the corresponding coupling scale by
$|\Delta g|_*$ gives
\begin{equation}
 |\Delta g|_*(F_{\rm loc})=
 \frac{|\Delta g|_*(F_{\rm loc}=1)}{\sqrt{F_{\rm loc}}}.
 \label{eq:density}
\end{equation}

The rescaled values retain the assumptions of Sec.~\ref{sec:reactors}:
small field variation during each flight, $m_aL_*\ll1$, adequate sampling
of the chosen field distribution, and the same treatment of reactor
distances and external solar information. The mass and sampling
conditions apply separately to each experiment.

\begin{figure}[t]
\centering
\includegraphics[width=\columnwidth]{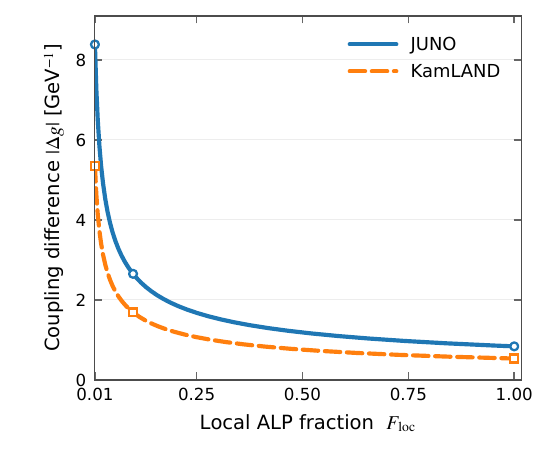}
\caption{\protect{Coupling scales that reproduce the
reference reactor damping as a function of the local ALP fraction
$F_{\rm loc}$, for $g_2=g_3$ and
$\rho_{\rm DM}=0.4\GeV\,\mathrm{cm}^{-3}$.
The solid JUNO and dashed KamLAND curves use the Gaussian-field values
in Table~\ref{tab:reactors}, rescaled using Eq.~\eqref{eq:density}
under the assumptions of Sec.~\ref{sec:reactors}.
Markers indicate local fractions of 1\%, 10\%, and 100\%.}}
\label{fig:reactor_abundance}
\end{figure}

Figure~\ref{fig:reactor_abundance} shows how the coupling scales for JUNO and KamLAND increase as the local ALP fraction decreases.
An ALP contribution of one percent of the local dark matter raises both
coupling scales by a factor of ten. The Gaussian values in
Table~\ref{tab:reactors} then become $8.38\GeV^{-1}$ for JUNO and
$5.34\GeV^{-1}$ for KamLAND. Changing the same local fraction for both
experiments leaves the ratio of their coupling scales unchanged.

The quantity probed by this comparison is thus
$|\Delta g|\sqrt{\rho_a}$. Separating the coupling from the local ALP
abundance requires independent information about the background or the
interaction.

\subsection{Longer terrestrial paths and matter}
\label{sec:matter}

The reactor measurements above probe electron-antineutrino survival. Muon-neutrino disappearance in accelerator and atmospheric experiments allows us to study the same ALP interaction over longer terrestrial paths.

The vacuum result in Eq.~\eqref{eq:L2} shows why these paths are of interest: at fixed ALP
density, the coupling difference needed for a given contrast decreases as
$1/L$ when the field changes little during each flight. For a Gaussian
field, a coupling difference $\Delta g$ reduces a vacuum interference
term by 10\% when
\begin{equation}
 \frac{(\Delta g)^2\rho_aL^2}{2}=-\ln(0.9).
 \label{eq:contrast}
\end{equation}
For the illustrative values $|\Delta g|=8.31\times10^{-3}\GeV^{-1}$
and $\rho_a=0.4\GeV\,\mathrm{cm}^{-3}$, this gives about 6200 km.
Longer flights also tighten the condition $m_aL\ll1$; for
$L=10^4$ km, it requires $m_a\ll2.0\times10^{-14}$ eV.
The observed contrast also depends on the field variations sampled during
data taking, as discussed in Sec.~\ref{sec:halo}.

Along these paths, matter also changes the neutrino evolution. Coherent
forward scattering on electrons adds a potential $\sqrt2G_FN_e(s)$
to the electron flavor, where $G_F$ is the Fermi constant, $N_e$ is the
electron number density, and $s$ is the distance along the path, from
production at $s=0$ to detection at $s=L$. The neutral-current potential
common to the three flavors contributes only an overall
phase~\cite{Coloma2018}. At leading ultrarelativistic order,
the three flavor amplitudes in $\psi_f$ obey
$\ii\,\dd\psi_f/\dd s=H_f(s)\psi_f$, with
\begin{equation}
 H_f(s)=\frac{UM^2U^\dagger}{2E}
 +\diag(\sqrt2G_FN_e(s),0,0)
 -\frac{\dd a}{\dd s}UgU^\dagger.
 \label{eq:matterH}
\end{equation}
Here $U$ is the neutrino mixing matrix, $E$ the neutrino energy,
$M^2=\diag(m_1^2,m_2^2,m_3^2)$, and
$g=\diag(g_1,g_2,g_3)$. The derivative $\dd a/\dd s$ follows the
neutrino trajectory and includes both temporal and spatial changes of
the field. For antineutrinos, $U$ is conjugated and the matter and ALP
potentials reverse sign.

The electron flavor is a superposition of vacuum mass states. The matter
term therefore generally mixes the components whose relative phases the
ALP changes. The two contributions must be included together when
solving Eq.~\eqref{eq:matterH}. For each field realization, this
Hamiltonian determines the amplitudes and hence the flavor probabilities.
We then average these probabilities over the field variations sampled
during data taking. The ALP potential remains independent of neutrino
energy at the order retained, while its combined action with the vacuum
and matter terms can produce an energy-dependent change in the flavor
probabilities.

To see how matter changes the ALP effect, consider an
atmospheric muon neutrino crossing $6200$ km of the Earth's mantle.
This is the distance obtained from the 10\% reduction of vacuum
interference in Eq.~\eqref{eq:contrast}. Approximating the mantle by
a constant density of $4.5\,\mathrm{g\,cm}^{-3}$, with $0.5$ electrons
per nucleon, gives an electron potential
$\sqrt{2}G_FN_e\simeq1.72\times10^{-13}\eV$.

The mixing angles and CP phase are those in
Appendix~\ref{app:decays}. For normal ordering, we take
$\Delta m_{21}^2=7.388\times10^{-5}\eV^2$ and
$\Delta m_{31}^2=2.509\times10^{-3}\eV^2$~\cite{JUNO207Days2026}.
The coupling difference and ALP density are the same as in the vacuum
estimate, with $g_2=g_3$. Since a common coupling contributes only an
overall phase, we set $g_1=0$ for this propagation calculation.

Taking $m_a=10^{-15}\eV$ gives $m_aL\simeq0.031$, so
the field derivative $\dd a/\dd s$ is nearly constant during each flight. Its value can still vary between events.
We assume that the events sample the Gaussian field distribution of Sec.~\ref{sec:halo}. The ALP potential difference between mass states 2 and 1 then has zero mean and standard deviation
$|\Delta g|\sqrt{\rho_a}\simeq1.46\times10^{-14}\eV$, at leading order in the halo velocity.

The Hamiltonian is then constant along each path. For each sampled value of the field derivative, the survival probability is
\[
 P_{\mu\mu}=\left|\left[e^{-\ii H_fL}\right]_{\mu\mu}\right|^2.
\]
Averaging these probabilities over the Gaussian distribution gives $\overline P_{\mu\mu}$.

The change produced by the ALP is
$\delta P_{\mu\mu}\equiv\overline P_{\mu\mu}-P_{\mu\mu}^{(0)}$,
where $P_{\mu\mu}^{(0)}$ is the probability without the ALP interaction
at the same matter density. Evaluating this difference at $3$, $5$,
and $8\GeV$ shows its energy dependence. For comparison, the vacuum
values use $N_e=0$ with all other inputs unchanged:
\[
 \begin{array}{c@{\qquad}rr}
 E\,[\mathrm{GeV}] & \delta P_{\mu\mu}^{\rm matter}
                   & \delta P_{\mu\mu}^{\rm vacuum}\\[3pt]
 3 & -9.65\times10^{-3} & -14.5\times10^{-3}\\
 5 & -1.23\times10^{-3} & -6.28\times10^{-3}\\
 8 & -0.28\times10^{-3} & -8.76\times10^{-3}
 \end{array}
\]

For this path and these energies, matter reduces the
magnitude of the ALP correction, most strongly at $8\GeV$. The 10\%
criterion in Eq.~\eqref{eq:contrast} refers to the affected vacuum
interference terms. The effect on the total survival probability also
depends on the weights and energy-dependent phases of those terms.

Existing accelerator and atmospheric searches constrain oscillation
damping for specified evolution laws
\cite{DeRomeri2023,Coloma2018,IceCube2024,ORCA2025}. For example,
Ref.~\cite{Coloma2018} defines dephasing in the basis that diagonalizes
the Hamiltonian in matter, whereas here $g$ is diagonal in the vacuum
mass basis. Applying these data to the derivative interaction requires
solving Eq.~\eqref{eq:matterH} with the electron density along each
path. The averaged probabilities are then combined with the incident
flux and detector response to obtain the predicted spectrum over the
energy range where the effective interaction applies.

\subsection{Astrophysical sources and cosmological abundance}
\label{sec:astro}

For astrophysical neutrinos, the distance dependence still follows from
the vacuum phase in Eq.~\eqref{eq:phase}, which is fixed by $a_D-a_P$.
When the field changes little during a flight, its change is approximately
the derivative along the trajectory multiplied by $L$, and the phase
variance grows as $L^2$. Over longer paths, this linear approximation no
longer holds: the endpoint difference $a_D-a_P$ need not continue to grow
with $L$.

For the zero-mean field of Sec.~\ref{sec:halo},
Eq.~\eqref{eq:covariance} relates the variance of $a_D-a_P$ to the
variance at each end of the path and the correlation $\avg{a_Da_P}$.
At fixed, finite variances in the source and detector environments, the
variance of their difference is bounded. If the two field values become
uncorrelated, the cross term vanishes and the variance becomes
$\avg{a_D^2}+\avg{a_P^2}$, or $2\avg{a^2}$ when the variances are
equal. A perfectly coherent sinusoidal mode can instead retain an
oscillatory dependence on distance. Keeping the two variances separate
also accounts for source and observer regions with different field
densities.

For this phase to affect a flavor probability, the mass components must
still interfere. Consider a state that enters a vacuum part of the path
as an incoherent mixture,
\[
 \rho_\nu=\sum_i p_i|\nu_i\rangle\langle\nu_i|,
 \qquad \sum_i p_i=1,
\]
where the nonnegative coefficients $p_i$ are the probabilities for the
individual mass states. Diagonal vacuum evolution multiplies each ket
and bra by opposite phases, which cancel within every term of
$\rho_\nu$. The populations $p_i$ and the detected flavor fractions
$\sum_i p_i|U_{\beta i}|^2$ consequently remain unchanged.

Flux measurements can still test processes that change the number of
arriving neutrinos. The SN1987A analysis of Ref.~\cite{TernesSN2025},
for example, allows neutrino loss along the trajectory and constrains
the resulting attenuation. The diagonal vacuum evolution above preserves
both the total flux and the mass-state populations. Sensitivity to the
derivative interaction can also arise during production in a dense
source or conversion in matter. In those regions, the signal depends on
how the ALP contribution combines with the source interactions and the
matter potential.

The same interaction can also be tested through a different observable
in the early Universe, where collisions rather than oscillation phases
become relevant. ALP production can change particle abundances and
scattering can redistribute neutrino momenta. Frequent scattering
impedes neutrino free streaming, and the cosmological effect depends on
the relevant rates compared with the expansion rate~\cite{HuangNath2018}.
These rates involve the individual couplings, particle masses, and
momentum distributions. A common shift $g_i\to g_i+g_0$ leaves the
leading oscillation phases unchanged but can change the collision
amplitudes. A collision calculation therefore requires coupling
information beyond the differences probed by oscillations.

For external Dirac neutrinos, the free equations of motion turn the
single-ALP vertex into a pseudoscalar coupling with dimensionless strength
$m_i g_i$ in our conventions. This follows from the mass contribution
to the current divergence in Appendix~\ref{app:propagation}. For
illustration, take $g_i=0.9\GeV^{-1}$ and $m_i=0.05$ eV. Then
$m_i g_i=4.5\times10^{-11}$. When the derivative interaction is removed
by a field redefinition, the transformed mass term also contains vertices
with two or more ALPs. Processes with several ALPs receive contributions
from these vertices as well as from the linear pseudoscalar coupling, at
the corresponding order~\cite{HuangNath2018,CriadoPerezVictoria2019}.

Structure formation provides a further test of the assumed background
abundance. In canonical ultralight-axion models, wave effects suppress
the growth of small-scale structure. The Lyman-$\alpha$ forest probes
the intervening matter distribution through hydrogen absorption in
quasar spectra. For axions making up all dark matter, the analysis of
Ref.~\cite{RogersPeiris2021} gives
$m_a>2\times10^{-20}$ eV at 95\% credibility.

Galaxy abundances as a function of ultraviolet luminosity provide
another measure of structure formation. The joint analysis of Hubble
Space Telescope galaxy luminosities and Planck cosmic microwave
background data in Ref.~\cite{Winch2025} constrains the cosmic axion
fraction to about 0.22 or less for $10^{-26}<m_a/\mathrm{eV}<10^{-23}$
and excludes an all-axion component below $10^{-21.6}$ eV, both at
95\% credibility. The fraction limit varies with mass and becomes
tighter within part of this interval. The inference uses a model of
halo formation and a relation between halo mass and galaxy luminosity.

Applying these abundance limits to oscillations requires a connection
between the cosmological mean density and the local Galactic field.
The cosmic fraction $f_{\rm cos}=\overline\rho_a/\overline\rho_{\rm DM}$
uses the mean densities denoted by the bars, whereas $F_{\rm loc}$ in
Eq.~\eqref{eq:density} uses the local densities. Identifying the two
assumes that ALPs and the other dark matter components cluster in the
same proportions in the local environment. For the fixed-amplitude mode, we assign the resulting local density to the sampled mode. Table~\ref{tab:cosmological_abundance} summarizes the
limits on the ALP mass and cosmic abundance discussed above.
\begin{table*}[t]
\caption{\protect{Abundance limits from structure
formation in canonical ultralight-axion models, all at 95\% credibility.
The subcomponent entry summarizes upper limits that vary with mass
and become tighter within part of the quoted interval.}}
\label{tab:cosmological_abundance}
\begin{ruledtabular}
\begin{tabular}{p{0.32\textwidth}p{0.22\textwidth}p{0.36\textwidth}}
Observation & ALP abundance & Limit\\
\hline
Lyman-$\alpha$ forest~\cite{RogersPeiris2021}
& All dark matter ($f_{\rm cos}=1$)
& $m_a>2\times10^{-20}$ eV\\[3pt]
HST galaxy UV luminosities and Planck~\cite{Winch2025}
& All dark matter ($f_{\rm cos}=1$)
& $m_a>10^{-21.6}$ eV\\[3pt]
HST galaxy UV luminosities and Planck~\cite{Winch2025}
& Subcomponent ($f_{\rm cos}<1$)
& $f_{\rm cos}\lesssim0.22$ for $10^{-26}<m_a/\mathrm{eV}<10^{-23}$\\
\end{tabular}
\end{ruledtabular}
\end{table*}

Within the illustrative reactor interval
$10^{-20}$--$10^{-13}$ eV, the Lyman-$\alpha$ result disallows
canonical all-axion dark matter below $2\times10^{-20}$ eV.
The upper limits on subcomponents from the luminosity analysis
lie at smaller masses, outside this interval. Moving to lower local
fractions requires larger coupling differences to obtain the same
contrast, as shown in Fig.~\ref{fig:reactor_abundance}. This raises
the question of whether couplings of that size are compatible with
independent laboratory tests.

\section{Electroweak embeddings and complementary probes}
\label{sec:ew}

The interaction responsible for the ALP-induced phase also allows real
ALPs to be emitted in decays. Which decay measurements constrain the same
coefficients depends on the other interactions that accompany the
neutrino current. Electroweak symmetry provides a concrete starting point:
neutrinos and charged leptons belong to the same doublets. We examine a
current of the entire doublet and an interaction that uses Higgs fields
to select neutrinos, following the decay signals that each produces.
Pion and kaon emission also test the neutrino-only low-energy interaction
with its usual charged weak current.

\subsection{The dimension-five doublet current}
\label{sec:d5}

The derivative interaction in Eq.~\eqref{eq:left} has dimension five. A
simple gauge-invariant realization at the same dimension couples the ALP to
the full left-handed lepton doublet,
\begin{equation}
 \mathcal L_5=(\partial_\mu a)\bar L_\alpha\gamma^\mu
 (G_L)_{\alpha\beta}L_\beta,
 \qquad G_L=G_L^\dagger .
 \label{eq:d5}
\end{equation}
Here $L_\alpha=(\nu_{\alpha L},\ell_{\alpha L})^T$, with
$\alpha=e,\mu,\tau$, contains the neutrino and charged lepton of each
family. The Hermitian coefficient matrix $G_L$ has mass dimension $-1$.
This shift-symmetric operator
belongs to the electroweak ALP effective theory
\cite{GrojeanKleyYao2023,BonillaEtAl2024}.

Because both members of the doublet appear in Eq.~\eqref{eq:d5}, the
interaction contains a neutrino current and a charged-lepton current with
the same flavor matrix. To see the consequence, choose a basis in which the
charged-lepton masses are diagonal. The neutrino flavor fields are then
related to the mass eigenstates by
$\nu_{\alpha L}=\sum_i U_{\alpha i}\nu_{iL}$. Matching the neutrino
current to the diagonal couplings used in the oscillation calculation gives
\begin{equation}
 G_L=U\diag(g_1,g_2,g_3)U^\dagger .
 \label{eq:d5alignment}
\end{equation}
If the three $g_i$ are equal, $G_L$ is proportional to the identity and
its charged-lepton current preserves flavor. The coupling differences
that generate the neutrino oscillation phase therefore also generate
off-diagonal charged-lepton couplings in general.

This connection is especially direct for the reactor benchmark $g_2=g_3$.
Using the unitarity of $U$, we find
$(G_L)_{e\mu}=-\Delta g_{21}U_{e1}U_{\mu1}^*$, which allows the
two-body decay $\mu\to ea$. When the ALP is effectively massless and the
electron mass is neglected, the width is
\begin{equation}
 \Gamma(\mu\to ea)=\frac{m_\mu^3}{32\pi}|(G_L)_{e\mu}|^2.
 \label{eq:muwidth}
\end{equation}
The operator couples to left-handed charged leptons, so the appropriate
comparison is the $V-A$ decay pattern discussed in
Ref.~\cite{CalibbiEtAl2021}. The TWIST search gives
$\mathrm{BR}(\mu\to ea)<5.8\times10^{-5}$ at 90\% confidence for this
pattern~\cite{TWIST2015,CalibbiEtAl2021}. It implies
$|(G_L)_{e\mu}|<1.22\times10^{-9}\GeV^{-1}$. Using the illustrative
mixing parameters listed in Appendix~\ref{app:decays}, which lie within
the ranges of Ref.~\cite{NuFIT2024}, we find
$|U_{e1}U_{\mu1}^*|\simeq0.249$. The TWIST limit then gives
$|\Delta g_{21}|\lesssim4.9\times10^{-9}\GeV^{-1}$. This is many
orders of magnitude below the reactor coupling scales in
Table~\ref{tab:reactors}. Unlike an oscillation measurement, the decay
probes the interaction without requiring an ALP background.

The muon decay constrains a particular combination of couplings. For
general diagonal $g$, define the two independent differences
$x=\Delta g_{21}=g_2-g_1$ and $y=\Delta g_{31}=g_3-g_1$. Unitarity of
$U$ then gives
\[
 (G_L)_{e\mu}=xU_{e2}U_{\mu2}^*+yU_{e3}U_{\mu3}^*.
\]
The decay width depends on the squared modulus of this sum. Its two terms can
interfere, with the extent of any cancellation determined by the mixing
matrix and the real coefficients $x$ and $y$. The numerical result above
uses $x=y$; other flavor patterns require their corresponding combination.

The experimental comparison also depends on the chirality of the
charged-lepton interaction. For example, an independent right-handed
current with coefficient matrix $G_R$,
$(\partial_\mu a)\bar\ell\gamma^\mu P_RG_R\ell$, where
$P_R=(1+\gamma^5)/2$, contributes to the same decay. In the limit
$m_e,m_a\to0$, the two chiral contributions add in the width,
\[
 \Gamma(\mu\to ea)=\frac{m_\mu^3}{32\pi}
 \left[|(G_L)_{e\mu}|^2+|(G_R)_{e\mu}|^2\right].
\]
They lead to different angular distributions for polarized muons, so a
limit obtained for the left-handed pattern must be interpreted with the
appropriate experimental acceptance~\cite{CalibbiEtAl2021}.

The diagonal charged-lepton couplings are also tested by laboratory and
stellar measurements~\cite{Giannottietal2017,Berlinetal2024}. Translating
those results involves the common part of the $g_i$ and any independent
right-handed charged-lepton current. The off-diagonal muon-decay coupling
depends only on the differences and gives a particularly strong test of
the chosen flavor pattern.

The charged-lepton partners also enter the field redefinition discussed in
Sec.~\ref{sec:ibp}. Rotating the whole doublet removes Eq.~\eqref{eq:d5}
without changing the gauge charged-current vertex, but transfers the ALP
dependence to lepton mass-generating interactions and potentially to
anomalous gauge couplings~\cite{BonillaEtAl2024,AdsheadLozanov2022}.
An oscillation prediction for this doublet theory must therefore include
the charged fields at production and detection as well. The muon bound
above follows directly from the charged-lepton coupling implied by
Eq.~\eqref{eq:d5alignment}. We next consider how Higgs fields can
distinguish the neutrino from its charged partner.

\subsection{Selecting neutrinos with Higgs fields}
\label{sec:d7}

The Higgs field offers a way to distinguish the neutrino from its charged
partner without breaking electroweak gauge invariance. After electroweak
symmetry breaking, its vacuum value selects a direction in the lepton
doublet. We can use that direction to construct an ALP interaction that
acts on neutrinos without the accompanying charged-lepton current of
Eq.~\eqref{eq:d5}.

Let $H$ be the Standard Model Higgs doublet and define
$\widetilde H=\ii\sigma^2H^*$, where $\sigma^2$ is the second Pauli matrix.
In unitary gauge, $H=(0,(v+h)/\sqrt2)^T$, with $v\simeq246\GeV$ the
Higgs vacuum expectation value and $h$ the physical Higgs field. It
follows that $\widetilde H^\dagger L_\alpha=(v+h)\nu_{\alpha L}/\sqrt2$:
the Higgs combination selects the neutrino component of the doublet.
Equivalently, the matrix acting on the doublet is
\begin{equation}
 \widetilde H\widetilde H^\dagger\longrightarrow
 \frac{(v+h)^2}{2}\begin{pmatrix}1&0\\0&0\end{pmatrix}.
 \label{eq:projector}
\end{equation}

This construction uses two Higgs fields, raising the dimension of the
derivative interaction from five to seven. The corresponding
gauge-invariant operator is
\begin{equation}
 \mathcal L_7=(\partial_\mu a)(\bar L_\alpha\widetilde H)\gamma^\mu
 (G_7)_{\alpha\beta}(\widetilde H^\dagger L_\beta),
 \label{eq:d7}
\end{equation}
where $G_7=G_7^\dagger$ has mass dimension $-3$. This is a particular
combination of the dimension-seven structures in the ALP effective
theory~\cite{GrojeanKleyYao2023}.

To connect this operator with the interaction used in the oscillation
calculation, expand the Higgs fields about their vacuum value and express
the neutrino current in the mass basis. With the same mixing matrix $U$
as in Sec.~\ref{sec:d5}, we obtain
\begin{equation}
 \begin{aligned}
 g&=\frac{v^2}{2}U^\dagger G_7U,\\
 \mathcal L_7&=(\partial_\mu a)\bar\nu\gamma^\mu P_Lg\nu
 \left(1+\frac{2h}{v}+\frac{h^2}{v^2}\right).
 \end{aligned}
 \label{eq:d7matching}
\end{equation}
The term independent of $h$ reproduces the low-energy neutrino
interaction in Eq.~\eqref{eq:left}. The remaining terms couple one or
two physical Higgs bosons to the same ALP--neutrino current. Their
strengths are fixed by $g$ in this single-operator construction.

In contrast with the dimension-five doublet current, this operator has
no direct ALP--charged-lepton bilinear at tree level. It nevertheless
predicts Higgs interactions, and the usual weak charged currents will
provide further tests at lower energies. Gauge invariance does not
require $g$ to be diagonal in the neutrino mass basis; that remains the
flavor assumption of our oscillation analysis. We first consider the
decay $h\to a\nu\bar\nu$, induced by the term linear in $h$.

\subsection{The invisible Higgs constraint}
\label{sec:higgs}

The term linear in $h$ in Eq.~\eqref{eq:d7matching} allows the Higgs to
produce an ALP together with a neutrino pair. Although the ALP couples
through a derivative, the decay amplitude survives in the limit of
massless neutrinos. Momentum conservation relates the momentum of the
emitted ALP to those of the Higgs and the two neutrinos. The free
neutrino equations of motion remove the contributions from the outgoing
neutrino momenta, while the Higgs momentum remains. The decay therefore
does not require a factor of $m_\nu$.

For $m_a,m_\nu\ll m_h$, summing over the final neutrino mass states and
integrating the three-body phase space gives
\begin{equation}
 \Gamma(h\to a\nu\bar\nu)=
 \frac{m_h^5}{1536\pi^3v^2}\Tr(g^\dagger g).
 \label{eq:hwidth}
\end{equation}
The amplitude and the phase-space integration are given in
Appendix~\ref{app:decays}. The trace is the sum of squared couplings to
all neutrino pairs, $\Tr(g^\dagger g)=\sum_{ij}|g_{ij}|^2$. For the real
diagonal matrix used in the oscillation calculation it becomes
$\sum_i g_i^2$. The same inclusive result holds in the massless limit for Majorana
neutrinos with the same left-handed current normalization. The Higgs
width thus depends on the strengths of the individual couplings,
whereas neutrino oscillations probe their differences.

When the ALP escapes detection together with the neutrinos, this channel
contributes to the invisible Higgs width. The ATLAS combination
sets an upper limit $B_{\rm inv}=0.107$ on the invisible branching ratio
at 95\% confidence~\cite{ATLASInvisible2023}. To relate that measurement
to $g$, we use the single dimension-seven operator of
Eq.~\eqref{eq:d7}, assuming that it remains valid at the Higgs mass and
that Higgs production follows the Standard Model rates used in the
experimental analysis. We take Eq.~\eqref{eq:hwidth} to describe the new
invisible width, $\Gamma_{\rm new}$. Including this contribution in the
total width, $\Gamma_h=\Gamma_h^{\rm SM}+\Gamma_{\rm new}$, gives
\begin{equation}
 \begin{aligned}
 \Gamma_{\rm new}&<\frac{B_{\rm inv}}{1-B_{\rm inv}}\Gamma_h^{\rm SM},\\
 \Tr(g^2)&<4.61\times10^{-5}\GeV^{-2},
 \end{aligned}
 \label{eq:hbound}
\end{equation}
where we used $m_h=125$ GeV, $v=246$ GeV and
$\Gamma_h^{\rm SM}=4.07$ MeV~\cite{HiggsWG}, neglecting the small
Standard Model invisible contribution.

To compare this bound with oscillation measurements, we must account for
the common coupling that oscillations cannot determine. Using the two
differences $x=\Delta g_{21}$ and $y=\Delta g_{31}$ defined in
Sec.~\ref{sec:d5}, the diagonal coefficients can be written as
$(g_1,g_2,g_3)=(c,c+x,c+y)$, with $c=g_1$ real. Changing $c$ leaves the leading
oscillation phases unchanged but changes the Higgs width through
$c^2+(c+x)^2+(c+y)^2$. For given $x$ and $y$, the smallest possible
width occurs at $c=-(x+y)/3$, where $g_1+g_2+g_3=0$. Hence
\begin{equation}
 \min_c\Tr(g^2)=\frac{2}{3}(x^2+y^2-xy).
 \label{eq:mintrace}
\end{equation}
This gives a lower bound on the invisible width at fixed coupling
differences, independent of the choice of the common coefficient.

\begin{figure}[htbp]
\centering\includegraphics[width=\columnwidth]{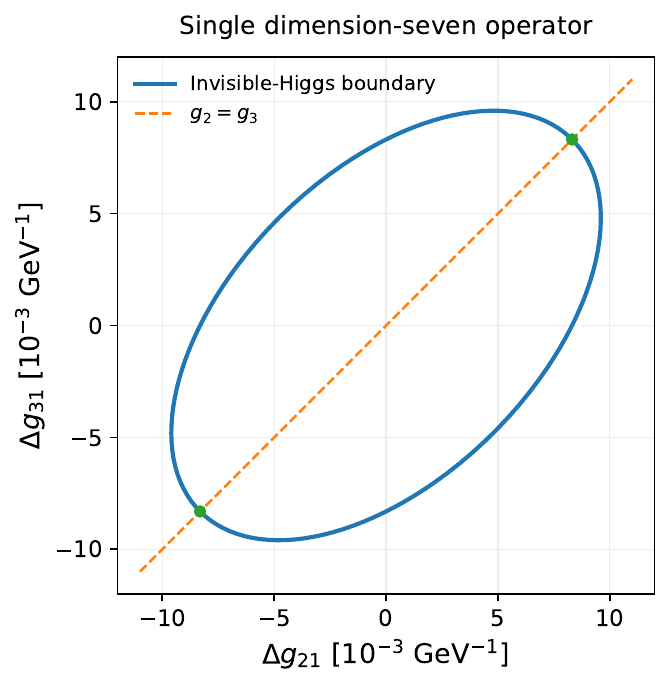}
\caption{Invisible-Higgs constraint on the coupling differences
$x=\Delta g_{21}$ and $y=\Delta g_{31}$, with both axes in units of
$10^{-3}\GeV^{-1}$. The ellipse is obtained by minimizing the decay
width over the common coefficient $c$. Its interior contains the
differences compatible with the ATLAS bound for at least one $c$.
The dashed line shows $g_2=g_3$, with intersections given by
Eq.~\eqref{eq:hdifference}. The calculation assumes a single
dimension-seven operator valid at $m_h$, Standard Model Higgs
production and invisible ALP final states.}
\label{fig:higgs}
\end{figure}

Combining Eqs.~\eqref{eq:hbound} and \eqref{eq:mintrace} gives the ellipse
in Fig.~\ref{fig:higgs}. For each point $(x,y)$ inside the ellipse,
there is at least one choice of $c$ that satisfies the Higgs limit.
Outside it, the smallest possible width already exceeds that limit.
The cross term $-xy$ makes the ellipse extend farther along $x=y$ than
along $x=-y$: differences with the same sign can give a smaller sum of
squared couplings. The dashed diagonal corresponds to the reactor
benchmark $g_2=g_3$, or $x=y\equiv\Delta g$. Its intersections with the
ellipse give
\begin{equation}
 |\Delta g_{21}|=|\Delta g_{31}|<8.31\times10^{-3}\GeV^{-1}.
 \label{eq:hdifference}
\end{equation}

In the dimension-seven scenario, the invisible-Higgs bound makes the ALP-induced damping at reactor baselines much smaller than the levels allowed by the reference fits. For $g_2=g_3$, reaching those levels would require local ALP fractions of $F_{\rm loc}\simeq1.0\times10^4$ for JUNO and $3.8\times10^3$ for KamLAND, using the fixed-amplitude mode and Eqs.~\eqref{eq:hdifference} and \eqref{eq:density}. Both values exceed the physical maximum $F_{\rm loc}=1$. Even if ALPs make up all of the local dark matter, the leading Gaussian dephasing exponent at the JUNO baseline is only $7.5\times10^{-6}$ at the Higgs limit. We next consider weak processes that probe the same neutrino--ALP interaction at lower energies.

\subsection{Neutral currents, charged currents, and meson emission}
\label{sec:charged}

The weak interactions that produce and detect neutrinos also provide ways
for a neutrino to emit an ALP. The outcome depends on the weak current
involved. A neutral current couples two neutrino fields, whereas a charged
current connects a neutrino to a charged lepton. This distinction becomes
particularly simple in the field representation of Sec.~\ref{sec:ibp},
where the ALP dependence has been moved from the neutrino kinetic term
to the interaction vertices.

Consider first the neutral current mediated by the $Z$ boson. For unitary
mixing among the three light neutrinos, this current is proportional to
the identity in the mass basis. The field rotation acts on both neutrino
legs and leaves the current unchanged. Equivalently, the tree-level
amplitudes for $Z\to\nu\bar\nu a$ with ALP emission from the neutrino
and antineutrino cancel when neutrino masses are neglected. The leading
unsuppressed emission amplitude therefore vanishes for this current;
mass corrections and other interactions are beyond this cancellation.

The charged current behaves differently because it joins a neutrino to
a charged lepton. In the neutrino-selective theory of
Eq.~\eqref{eq:d7matching}, only the neutrino carries the ALP-dependent
rotation. Expanding the transformed weak vertex then produces a contact
interaction among the ALP, the $W$ boson, a charged lepton, and a neutrino.
Its coupling to a charged-lepton flavor $\ell$ and neutrino mass state $i$
is proportional to $(Ug)_{\ell i}$. Because the neutrino mass state is not
identified in these decays, the probabilities for the distinct final states
are added. For the diagonal $g$ used in our oscillation calculation, this
gives
\begin{equation}
 G_\ell^2=(Ug^2U^\dagger)_{\ell\ell}
         =\sum_i|U_{\ell i}|^2g_i^2,
 \qquad \sum_\ell G_\ell^2=\Tr(g^2).
 \label{eq:row}
\end{equation}
Here $G_\ell\geq0$ is the square root of the sum, with mass dimension
$-1$. It measures the total strength of ALP emission associated with a
given charged lepton. For negligible final masses, the corresponding
$W$ decay satisfies
\begin{equation}
 \frac{\Gamma(W\to\ell\nu a)}{\Gamma(W\to e\nu)}
 =\frac{3m_W^2}{256\pi^2}G_\ell^2,
 \label{eq:W}
\end{equation}
in agreement with the neutrino-only case of
Ref.~\cite{AltmannshoferDrorGori2023}. Taking $m_W=80.4$ GeV and
$\Gamma(W\to e\nu)=0.226$ GeV, we find an inclusive new width
$1.735\GeV^3\Tr(g^2)$. Using the Higgs restriction in
Eq.~\eqref{eq:hbound}, this is below $8.1\times10^{-5}$ GeV. The
total-width constraint discussed in Ref.~\cite{AltmannshoferDrorGori2023}
is therefore weaker for the same operator assumed valid at these
energies.

Pions and kaons allow us to test the charged-current interaction at much
lower energies, where the $W$ is replaced by the usual four-fermion weak
interaction. Let $P^\pm$ be a charged pseudoscalar meson, either
$\pi^\pm$ or $K^\pm$, decaying with $\ell=e$ or $\mu$. For $P^-$,
the quark current is described by the decay constant $f_P$, defined by
$\langle0|\bar q\gamma^\mu\gamma^5Q|P^-(p_P)\rangle=\ii f_Pp_P^\mu$.
Here $p_P$ is the meson four-momentum, and $q,Q$ label the quark
flavors. The relevant Cabibbo--Kobayashi--Maskawa element $V_{qQ}$ is
$V_{ud}$ for pions and $V_{us}$ for kaons. Charge conjugation gives the
corresponding positive-meson rate.

The ordinary two-body decay $P\to\ell\nu$ is suppressed by the
charged-lepton mass. With a massless neutrino, the meson momentum equals the sum
of the lepton and neutrino momenta; using their equations of motion makes
the amplitude proportional to $m_\ell$. This is the usual helicity
suppression of a leptonic pseudoscalar decay. When an ALP is emitted, momentum conservation instead gives
$p_P=p_\ell+p_\nu+p_a$. The extra term involving the ALP momentum can
survive when $m_\ell$ tends to zero, so the neutrino-only contact produces
an unsuppressed three-body contribution. Summing the neutrino final states
and integrating the phase space gives
\begin{equation}
 \Gamma(P\to\ell\nu a)=
 \frac{G_F^2|V_{qQ}|^2f_P^2m_P^5}{3072\pi^3}G_\ell^2
 F\!\left(\frac{m_\ell^2}{m_P^2},\frac{m_a^2}{m_P^2}\right).
 \label{eq:meson}
\end{equation}
Here $G_F$ is the Fermi constant, $m_P$ and $m_a$ are the meson and ALP
masses, and $F$ accounts for the masses in both the amplitude and the
three-body phase space. It is normalized by $F(0,0)=1$ and derived in
Appendix~\ref{app:decays}. The notation $\nu$ includes the neutrino or
antineutrino appropriate to the charge of $P$. The result applies to the
neutrino-only interaction; the dimension-five doublet current also has
ALP emission from the charged lepton, which changes this amplitude
\cite{AltmannshoferDrorGori2023,ExoticPions2026}.

For $m_a=0$, the numerical inputs given in Appendix~\ref{app:decays}
yield
$\mathrm{BR}(\pi\to e\nu a)=4.81\times10^{-5}\GeV^2G_e^2$
and
$\mathrm{BR}(K\to\mu\nu a)=9.09\times10^{-4}\GeV^2G_\mu^2$.
These rates depend on the individual $g_i$ through Eq.~\eqref{eq:row}, whereas the reactor comparison depends only on their differences.
To relate the pion rate to the reactor scales, we use $g_2=g_3$, with
$\Delta g=g_2-g_1=g_3-g_1$. The common part of the couplings does not
affect the oscillation phase, but it enters the decay rate. Minimizing
the pion rate over this common coefficient gives
$G_e^2\geq w_e(1-w_e)(\Delta g)^2$, where $w_e=|U_{e1}|^2$.
With the mixing parameters in Appendix~\ref{app:decays}, the minimum
pion branching fraction is
$1.05\times10^{-5}(|\Delta g|\,\mathrm{GeV})^2$.
At the single-mode scales in Table~\ref{tab:reactors}, this gives
$7.2\times10^{-6}$ for JUNO and $2.7\times10^{-6}$ for KamLAND.

PIENU searched for $\pi^+\to e^+\nu X$, with $X$ invisible,
using the positron energy spectrum~\cite{PIENU2021}.
To compare our prediction with the ALP benchmark studied in
Ref.~\cite{ExoticPions2026}, we use $x=2E_e/m_\pi$ for the
positron energy in the pion rest frame. Neglecting final-state
masses, Eq.~\eqref{eq:Fmeson} gives the normalized spectrum
\[
 \frac{1}{\Gamma}\frac{\dd\Gamma}{\dd x}=12x^2(1-x),
 \qquad 0\leq x\leq1.
\]
The benchmark in that study couples the ALP to the charged lepton
rather than to the neutrino, but gives the same normalized positron
spectrum in this limit. The branching-fraction limit shown in its
Fig.~7 is of order $10^{-6}$ for $m_a\ll m_\pi$.
Comparing this value with the minimum pion rate above gives an
approximate PIENU sensitivity of $|\Delta g|\sim0.3\GeV^{-1}$
for an invisible ALP and $g_2=g_3$. This is below both single-mode
reactor scales in Table~\ref{tab:reactors}.

The kaon search uses the missing mass reconstructed from the observed
charged lepton. For a kaon decay, the quantity
$q_{\rm miss}^2=(p_K-p_\ell)^2=(p_\nu+p_a)^2$ is the invariant mass
squared of the neutrino--ALP pair. The two invisible particles can have
nonzero invariant mass even when each is massless, so
$K\to\ell\nu a$ gives a continuous distribution. In the ordinary
two-body decay $K\to\ell\nu$, the true missing mass is instead zero
for a negligible neutrino mass.

\begin{figure}[htbp]
\centering\includegraphics[width=\columnwidth]{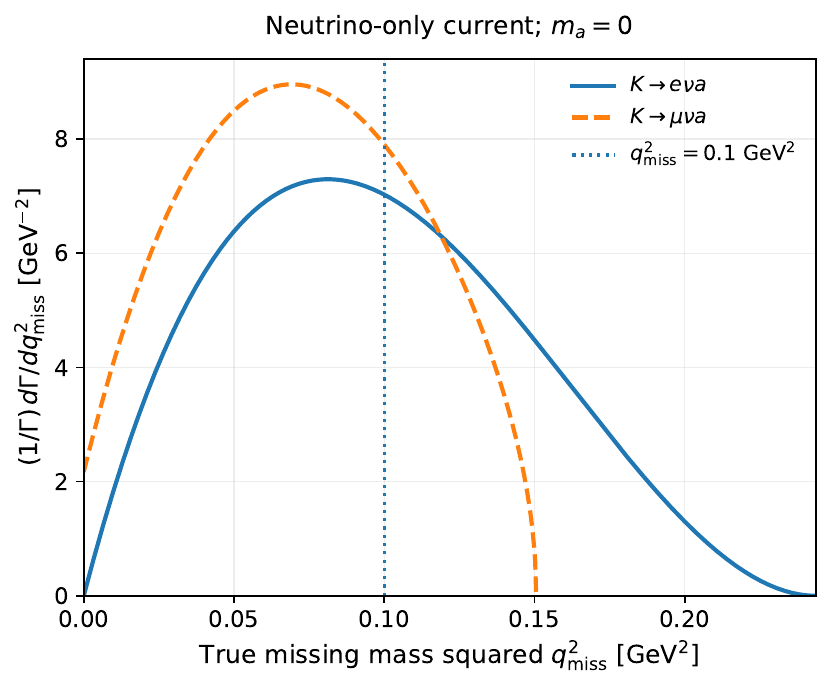}
\caption{Normalized true missing-mass distributions for
$K\to e\nu a$ (solid) and $K\to\mu\nu a$ (dashed), with negligible
neutrino and ALP masses. The horizontal axis is
$q_{\rm miss}^2=(p_K-p_\ell)^2$ and the vertical axis is
$(1/\Gamma)\dd\Gamma/\dd q_{\rm miss}^2$ in $\GeV^{-2}$;
each curve has unit area. The dotted line marks
$q_{\rm miss}^2=0.1\GeV^2$, the lower threshold of the NA62
muon-channel search region~\cite{NA622021}. The fraction of the
muon distribution above it is $0.266$ before detector effects.
The spectra follow Eq.~\eqref{eq:Fmeson}.}
\label{fig:mesons}
\end{figure}

Figure~\ref{fig:mesons} shows the spectra for the electronic and muonic
kaon channels, each divided by its own partial width. Their areas are
therefore one, and their heights describe the distribution of events
rather than the relative branching fractions. The muon mass shortens
the allowed interval to $q_{\rm miss}^2\le(m_K-m_\mu)^2$ and also changes
the shape through the decay amplitude and phase space. The figure is
calculated before detector resolution and event selection.

The threshold shown in the figure is used in the NA62 search for a muon
and invisible particles~\cite{NA622021}. In this selected region, the
collaboration reports a 90\%-confidence upper limit of 1184 signal events,
using an effective normalization of $N_K=1.14\times10^{10}$ kaon decays.
We retain this counting limit and compare it with the yield from our
spectrum using Eq.~\eqref{eq:acceptance}. The effective normalization
already includes losses common to the signal and the normalization
channel; its relation to the remaining acceptance is described in
Appendix~\ref{app:decays}.

For our muon spectrum, the fraction above the true threshold
$q_{\rm miss}^2=0.1\GeV^2$ is $f_{\rm cut}=0.26556$. Neglecting
migration across this threshold, the selected fraction can be written
as $f_{\rm cut}\bar\epsilon$, where $\bar\epsilon$ is the mean
acceptance of the remaining selections for this spectrum in the NA62
normalization. The published counting limit then gives
\[
 G_\mu<\frac{0.0207}{\sqrt{\bar\epsilon}}\GeV^{-1}.
\]

For $g_2=g_3$, the dependence on the common coupling can again be
removed by minimization. With $w_\mu=|U_{\mu1}|^2$, the result is
$G_\mu^2\geq w_\mu(1-w_\mu)(\Delta g)^2$, and hence
\[
 |\Delta g|<\frac{0.0207}
 {\sqrt{\bar\epsilon\,w_\mu(1-w_\mu)}}\GeV^{-1}.
\]
The mixing inputs in Appendix~\ref{app:decays} give
$w_\mu\simeq0.0915$ and a coefficient of
$0.072/\sqrt{\bar\epsilon}\GeV^{-1}$. The acceptance
$\bar\epsilon$ has not been determined for our neutrino--ALP spectrum. The expressions above therefore give a
conditional comparison with the NA62 selected-event limit. A dedicated
bound would require the detector acceptance and reconstructed
missing-mass distribution for this signal, together with the
corresponding background treatment. The meson calculation provides the
ingredients for this comparison without requiring the low-energy
interaction to remain valid at the Higgs mass.

\subsection{What the complementary constraints establish}
\label{sec:validity}

The decay calculations identify which combinations of the low-energy
coefficients are tested once their accompanying interactions are
specified. Table~\ref{tab:ew} collects the results for the two
electroweak operators. It also includes the meson tests that follow
already from the neutrino-only interaction and its weak vertices.

\begin{table*}[t]
\caption{Processes associated with the two electroweak operator choices at
leading order. The comparisons in $|\Delta g|$ use $g_2=g_3$, $\Delta g=g_2-g_1=g_3-g_1$, and the mixing inputs of Appendix~\ref{app:decays}. The muon limit is at 90\% confidence. The Higgs limit is at 95\% and assumes a single operator valid at $m_h$, Standard Model Higgs production, and an invisible ALP. The pion and kaon comparisons use an invisible, effectively massless ALP. The pion entry is an approximate sensitivity inferred by matching the leading positron spectrum to that of the PIENU reanalysis. The kaon entry converts the NA62 90\%-confidence event count, retaining the signal acceptance $\bar\epsilon$ defined in Sec.~\ref{sec:charged}}.
\label{tab:ew}
\begin{ruledtabular}
\begin{tabular}{p{0.17\textwidth}p{0.32\textwidth}p{0.41\textwidth}}
Probe & Dimension-five doublet current & Neutrino-selective dimension-seven operator\\
\hline
$\mu\to ea$ & Charged-lepton flavor-changing current; $|\Delta g|\lesssim4.9\times10^{-9}\GeV^{-1}$ with the stated mixing & No direct charged-lepton current at tree level; radiative effects depend on matching\\[2pt]
$h\to a\nu\bar\nu$ & Neutrino emission is mass suppressed; the Higgs contact in Eq.~\eqref{eq:d7matching} is absent & Invisible decay; $|\Delta g|<8.31\times10^{-3}\GeV^{-1}$ after minimizing the common coupling\\[2pt]
$Z\to\nu\bar\nu a$ & Neutrino-leg emission cancels for $m_\nu=0$ & The same leading cancellation\\[2pt]
$W\to\ell\nu a$ & Doublet rotation preserves the weak vertex; charged-lepton mass interactions contribute & Neutrino-selective contact, Eq.~\eqref{eq:W}\\[2pt]
$\pi\to e\nu a$ & Emission from both lepton partners must be included & PIENU comparison gives $|\Delta g|\sim0.3\GeV^{-1}$\\[2pt]
$K\to\mu\nu a$ & Emission from both lepton partners must be included &
NA62 count gives $|\Delta g|\lesssim0.072/\sqrt{\bar\epsilon}\GeV^{-1}$
at fixed signal acceptance\\[2pt]
Oscillations & Production and detection involve the correlated charged-lepton interactions & The low-energy neutrino phase is Eq.~\eqref{eq:phase}; the same coefficients enter Higgs and charged-current decays
\end{tabular}
\end{ruledtabular}
\end{table*}

The consequences differ markedly between the two embeddings. For the
dimension-five doublet current, the nonuniversal neutrino coupling is
accompanied by charged-lepton flavor violation, and the muon-decay bound
is much stronger than the reactor scale in the chosen flavor pattern.
The dimension-seven operator selects neutrinos at tree level, while its
Higgs vertices constrain the same pattern to
$|\Delta g|<8.31\times10^{-3}\GeV^{-1}$ when that operator is valid at
the Higgs mass. The bounds restrict the same coefficients that enter the low-energy
neutrino current. In the doublet theory, their oscillation prediction
also requires the correlated charged fields at production and detection;
the neutrino-selective operator reproduces the phase calculated above.

At lower energies, pion and kaon decays provide complementary tests of
the neutrino-only interaction. The rates and spectra derived in
Sec.~\ref{sec:charged} allow us to compare the predicted signals with
PIENU and NA62. For $g_2=g_3$, the approximate PIENU sensitivity reaches
smaller coupling differences than the reactor matching scales, while the NA62 comparison still depends on the signal acceptance. The Higgs and meson rates depend on different combinations of the individual $g_i$. At fixed $\Delta g$, the common coupling that minimizes one decay rate need not minimize the others. A combined analysis must therefore use the same $g_i$ in all channels.

The energy range of an effective operator determines which comparisons
can be made. For example, a difference of order the JUNO single-mode scale,
$|\Delta g|\simeq0.83\GeV^{-1}$, corresponds in the dimension-seven
coefficient to $2|\Delta g|/v^2\simeq2.7\times10^{-5}\GeV^{-3}$.
Expressed as a single inverse-cubic scale with a unit coefficient,
this is about $33\GeV$. It characterizes the size of the operator
coefficient; the physical cutoff depends on the underlying theory. This coefficient
alone does not establish a weakly coupled description valid well above
the electroweak scale. Applying the operator at $m_h$ requires the
validity assumption used in the Higgs comparison. The smallness of the ALP potential relative to the neutrino
energy, used in the propagation calculation, is a separate condition.
Additional operators and quantum corrections can also change the
relations among neutrino, Higgs, and charged-lepton couplings
\cite{BonillaEtAl2024,AdsheadLozanov2022}.

Oscillations and decays thus provide complementary information about
the interaction. The phase calculation determines how an ALP background
changes neutrino interference through $|\Delta g_{ij}|\sqrt{\rho_a}$.
Decay measurements constrain other combinations of the $g_i$ without
requiring a halo population. Specifying the interactions that relate
them connects the predicted reduction of contrast to independent tests
of the coupling strengths that can produce it.

\section{Discussion and conclusions}
\label{sec:conclusions}

An ultralight ALP background can reduce the measured contrast of neutrino
oscillations when events sample different unresolved field phases. For the
neutrino-only derivative interaction diagonal in the mass basis, the
leading ALP contribution to the relative phase in vacuum is
$-\Delta g_{ij}(a_D-a_P)$ in the ultrarelativistic limit. It depends only
on the field at production and detection. We obtain this phase either by
integrating the energy shift along the trajectory or by redefining the
neutrino fields and transforming the weak production and detection
vertices. Evolution remains unitary in each field realization and
preserves the mass-state populations. The reduction of contrast arises
from averaging probabilities for events with different relative phases.

The distribution of those phases determines the interference factor.
Uniform sampling of the phase of a single sinusoid gives $J_0(A_{ij})$,
while a zero-mean Gaussian field gives an exponential set by the phase
variance. When the field changes little during a flight, the phase spread
scales as $|\Delta g_{ij}|\sqrt{\rho_a}L$. For a small spread, both
descriptions give the same leading reduction of contrast, proportional
to $L^2$ and independent of neutrino energy. The two descriptions differ
at higher orders. The ordinary vacuum phase retains its energy dependence.
For observations of finite duration, event timing determines how the
phases are sampled. An additional phase that stays nearly fixed during
data taking shifts the oscillation pattern; sampled variations can
reduce its contrast.

This energy behavior follows from the derivative interaction. A scalar
that modulates the neutrino mass matrix acts through $M^2/(2E)$ and can
instead give an $L^2/E^2$ suppression
\cite{Berlin2016,DevMachadoMartinez2021,LosadaEtAl2022,
Chathirathas2026,Airoldi2026}. The flavor structure also matters:
off-diagonal derivative couplings can transfer population between mass
states. For smooth, stationary fluctuations with decaying correlations,
we obtain the same transition rate in the original and transformed field
descriptions. This result applies when transition probabilities are small
and the flight time is much longer than the field correlation time.
The rate contains an explicit $E^{-2}$ factor from the squared vacuum
energy splitting, with further energy dependence from the field spectrum
evaluated at that splitting.

Using published reactor damping limits at representative baselines,
we obtain coupling scales of $0.830$--$0.838\GeV^{-1}$ for JUNO
and $0.510$--$0.534\GeV^{-1}$ for KamLAND. These values assume
$g_2=g_3$, $|\Delta g|=|\Delta g_{21}|=|\Delta g_{31}|$, and
$\rho_a=0.4\GeV\,\mathrm{cm}^{-3}$; the ranges reflect the two field
distributions. The scales vary as $\rho_a^{-1/2}$ and are nearly
independent of the ALP mass when the field changes little during each
flight and the observations adequately sample the assumed distribution.
An illustrative common mass interval is
$10^{-20}\lesssim m_a/\mathrm{eV}\lesssim10^{-13}$ under these
conditions. KamLAND's longer path gives a smaller coupling scale in this
comparison. For the alternative pattern $g_1=g_2$, only the fast
interference terms are affected, giving JUNO coupling scales about four to five times larger than for $g_2=g_3$.
The conversion retains the external solar information used in the published fits. A fit tailored to the ALP model would use the individual reactor distances and event times and treat solar propagation
within the same model.

The vacuum scaling motivates searches for smaller coupling differences
over longer terrestrial paths. In matter, the vacuum, matter, and ALP
terms must be included in the same Hamiltonian, with the field average
taken after propagation. In the atmospheric example, matter changes the size and energy dependence of the ALP correction to the muon-neutrino survival probability. At astrophysical distances, the finite field variances at production and detection bound the phase variance. A mixture that is already incoherent in the mass basis remains unchanged under
diagonal evolution in vacuum.

Cosmological limits from particle collisions depend on the individual couplings and particle distributions, while structure formation constrains the cosmic ALP abundance. Relating those abundance limits to
oscillations requires a model for how ALPs cluster relative to the other dark matter components, since the propagation calculation uses the local
density.

The electroweak operators connect the neutrino couplings to independent
decay measurements. The dimension-five doublet current links nonuniversal
neutrino couplings to charged-lepton flavor violation. The accompanying
charged-lepton interactions also enter neutrino production and detection.
The dimension-seven operator selects neutrinos at tree level and fixes
their interactions with the Higgs. For $g_2=g_3$ and a single such
operator valid at $m_h$, minimizing over the common coupling gives
$|\Delta g|<8.31\times10^{-3}\GeV^{-1}$. This Higgs bound assumes
Standard Model production rates and an invisible ALP.
The muon and Higgs limits lie well below the reactor scales in their
respective realizations.

Pion and kaon decays test the neutrino-only interaction at much lower
energies. For an invisible, effectively massless ALP and $g_2=g_3$,
matching the leading positron spectrum to the PIENU reanalysis gives
an approximate sensitivity of $|\Delta g|\sim0.3\GeV^{-1}$, below
the single-mode reactor scales. The NA62 selected-event limit gives
$|\Delta g|\lesssim0.072/\sqrt{\bar\epsilon}\GeV^{-1}$, where
$\bar\epsilon$ is the mean acceptance for the remaining selections
above the true missing-mass threshold. This estimate neglects migration
across the threshold, and its interpretation depends on the acceptance
for the neutrino--ALP spectrum.

Neutrino oscillations probe how a classical ALP background changes
interference, while rare decays constrain the underlying couplings
without requiring a halo population. The relation between these tests
depends on the interactions accompanying the neutrino current and on
the energy range in which the effective theory is valid.

\acknowledgments
B.~A.~Couto e Silva and B.~L.~S\'anchez-Vega acknowledge financial support from the Funda\c{c}\~ao de Amparo \`a Pesquisa do Estado de Minas Gerais (FAPEMIG), with B.~L.~S\'anchez-Vega specifically supported under Grant No.~APQ-02289-26. B.~A.~Couto e Silva acknowledges additional support from the Coordena\c{c}\~ao de Aperfei\c{c}oamento de Pessoal de N\'ivel Superior (CAPES). During the preparation of this manuscript, the authors used OpenAI's ChatGPT as a research-assistance tool for numerical auditing, figure preparation, and editorial revision. The authors critically reviewed and validated all AI-assisted material and remain fully responsible for the accuracy, integrity, originality, and scientific content of the manuscript.

\appendix

\section{Currents, mass corrections, and field representations}
\label{app:propagation}

We derive the current identity behind the phase calculation and check
the mass corrections. We then compare the two field representations
over one neutrino flight and derive the perturbative transition rate.

\subsection{Current divergence with weak sources}

For diagonal $g$, integrating Eq.~\eqref{eq:left} by parts gives
$-g_i a\partial_\mu J_i^\mu$, up to a total derivative, with
$J_i^\mu=\bar\nu_i\gamma^\mu P_L\nu_i$. We use Dirac masses and work
at tree level. For the semileptonic weak interaction, write
$\mathcal L_{\rm CC}=\sum_i(\mathcal O_i+\mathcal O_i^\dagger)$,
where $\mathcal O_i=K_\mu\bar\ell_{\alpha L}\gamma^\mu
U_{\alpha i}\nu_{iL}$. The flavor $\alpha$ is summed, and $K_\mu$
contains the weak coefficient and hadronic current.

Under $\delta\nu_{iL}=\ii\epsilon_i(x)\nu_{iL}$, with the other
fields fixed, the kinetic term varies by
$-(\partial_\mu\epsilon_i)J_i^\mu$, the mass term by
$\ii\epsilon_i m_i\bar\nu_i\gamma^5\nu_i$, and the weak interaction
by $\ii\epsilon_i(\mathcal O_i-\mathcal O_i^\dagger)$.
Integrating the first term by parts and using the equations of motion
gives
\begin{equation}
 \partial_\mu J_i^\mu=-\ii m_i\bar\nu_i\gamma^5\nu_i
 +\ii(\mathcal O_i^\dagger-\mathcal O_i).
 \label{eq:currentdiv}
\end{equation}
The mass mixes the two chiralities, while the weak term creates or
absorbs a neutrino and survives when $m_i=0$. Using the free neutrino
equations would omit this weak contribution.

The same terms follow from the field redefinition
\begin{equation}
 \nu_i=e^{\ii g_i aP_L}\chi_i,
 \qquad
 \bar\nu_i=\bar\chi_i e^{-\ii g_i aP_R}.
 \label{eq:fullrotation}
\end{equation}
Here $P_{L,R}=(1\mp\gamma^5)/2$. The derivative acting on the
left-handed phase cancels Eq.~\eqref{eq:left}. The mass term becomes
\begin{equation}
 -m_i\bar\nu_i\nu_i
 \longrightarrow
 -m_i\bar\chi_i e^{-\ii g_i a\gamma^5}\chi_i,
 \label{eq:massrotation}
\end{equation}
while $\mathcal O_i$ and $\mathcal O_i^\dagger$, now evaluated with
$\chi_{iL}$, acquire $e^{\ii g_i a}$ and $e^{-\ii g_i a}$,
respectively. Expanding the transformed mass and weak terms to first
order in $g_i a$ reproduces $-g_i a\partial_\mu J_i^\mu$ using
Eq.~\eqref{eq:currentdiv}.
Vertices with two or more ALPs must be retained at the corresponding
perturbative order~\cite{CriadoPerezVictoria2019}. At loop level, the
change of the path-integral measure can also generate local
terms~\cite{AdsheadLozanov2022}.

At leading ultrarelativistic order, production and detection of
neutrino component $i$ carry $e^{-\ii g_i a_P}$ and
$e^{\ii g_i a_D}$. Their product gives
$e^{\ii g_i(a_D-a_P)}$ in Eq.~\eqref{eq:amplitude}, matching the
phase obtained from the original propagation.

\subsection{Helicity projection and Dirac--Majorana comparison}

Consider a negative-helicity Dirac spinor with momentum
$\mathbf p=p\widehat{\mathbf n}$, mass $m$, energy $E=\sqrt{p^2+m^2}$,
speed $v_\nu=p/E$, and normalization $u^\dagger u=2E$.
Its left- and right-handed norms are $E+p$ and $E-p$.
Since its spin points opposite to $\widehat{\mathbf n}$, the
projections with $\alpha^k=\gamma^0\gamma^k$ are
\[
 \frac{u^\dagger P_Lu}{2E}=\frac{1+v_\nu}{2},\qquad
 \frac{u^\dagger\bm\alpha P_Lu}{2E}
 =\frac{1+v_\nu}{2}\widehat{\mathbf n}.
\]
The expectation value of
$-g[(\partial_ta)P_L+(\bm\alpha\cdot\nabla a)P_L]$
then gives Eq.~\eqref{eq:shift}. For $v_\nu\simeq1$,
$\partial_ta+\widehat{\mathbf n}\cdot\nabla a$ is the derivative
along the trajectory, yielding Eq.~\eqref{eq:phase}.

For the axial interaction
$(c/f)(\partial_\mu a)\bar\nu\gamma^\mu\gamma^5\nu$, the same
projection gives
$(c/f)(v_\nu\partial_ta+\widehat{\mathbf n}\cdot\nabla a)$.
Thus $g=-c/f$ matches the leading relativistic potential, while the
mass corrections depend on the current.

To compare Dirac and Majorana masses, take a homogeneous field $a(t)$,
which conserves momentum and helicity. The Dirac amplitudes are the
left- and right-handed components. The Majorana amplitudes are a
canonically normalized Weyl field $n$ and its charge conjugate, with
mass term $-(mnn+\mathrm{h.c.})/2$~\cite{DreinerHaberMartin2010}.
Defining $b=g\dot a$ with the same left-handed current normalization,
the negative-helicity Hamiltonians are
\begin{equation}
 H^D_\nu=\begin{pmatrix}p-b&m\\m&-p\end{pmatrix},\qquad
 H^M_\nu=\begin{pmatrix}p-b&m\\m&-p+b\end{pmatrix}.
 \label{eq:modes}
\end{equation}
The Dirac right-handed component has no ALP potential; the Majorana
conjugate has the opposite potential.

The rotations $\diag(e^{\ii ga},1)$ and
$\diag(e^{\ii ga},e^{-\ii ga})$ cancel these background terms.
The upper-right mass entries become $me^{-\ii ga}$ and
$me^{-2\ii ga}$, respectively, with conjugate lower-left entries.
For Majorana, $n=e^{\ii ga}\xi$ equivalently changes the Weyl mass
coefficient to $me^{2\ii ga}$.

For constant $b$, the eigenvalues of Eq.~\eqref{eq:modes} connected
to the positive free energy are
\[
 \begin{aligned}
 E_+^D&=-\frac b2+\sqrt{(p-b/2)^2+m^2},\\
 E_+^M&=\sqrt{(p-b)^2+m^2}.
 \end{aligned}
\]
Their first-order shifts are $-b(1+v_\nu)/2$ and $-bv_\nu$.
Expanding $v_\nu=\sqrt{1-m^2/E^2}$ gives
\begin{equation}
 \begin{aligned}
 \delta E^D&=-b\left(1-\frac{m^2}{4E^2}+\cdots\right),\\
 \delta E^M&=-b\left(1-\frac{m^2}{2E^2}+\cdots\right).
 \end{aligned}
 \label{eq:DM}
\end{equation}
Both give $-b$ at leading order. The leading phase reverses sign
for antineutrinos. A symmetric distribution of $a_D-a_P$ then gives
the same real vacuum interference factor $\mathcal C_{ij}$.

To check the phase accumulated over a flight of duration $T$, keep
the homogeneous field and terms linear in $b_i(t)=g_i\dot a(t)$.
Assume $|b_i|\ll E_i$ and slow
variation that allows the state to follow its positive-energy
solution. Let $E_i=\sqrt{p^2+m_i^2}$, $v_i=p/E_i$, and
$\Delta a=a(T)-a(0)=a_D-a_P$, with common production and detection
times for the interfering components. The Dirac result is
\begin{equation}
 \begin{aligned}
 \Phi_i^{(a)}&=-g_i\Delta a+\delta\phi_i,\\
 \delta\phi_i&=\frac{1-v_i}{2}g_i\Delta a,\qquad
 |\delta\phi_i|\le\frac{1-v_i}{2}\int_0^T|b_i(t)|\dd t.
 \end{aligned}
 \label{eq:phaseerror}
\end{equation}
For Majorana, the correction has $1-v_i$ in place of $(1-v_i)/2$.
The phase approximation requires
$|\delta\phi_i-\delta\phi_j|\ll1$. Higher-order background terms
and corrections to adiabatic evolution must also remain small after
integration. Since $v_i$ depends on the mass, a common part of the
$g_i$ can enter these subleading relative phases.

\subsection{Finite-time relation to the commutator interaction}
\label{app:finite}

We first compare evolution over one flight, then extract the part of
an averaged transition probability that grows linearly with its
duration $T$. This flight time is distinct from the observation time
that controls phase sampling in Sec.~\ref{sec:halo}.

At leading ultrarelativistic order, use $H_\nu=H_0-g\dot a$, with
$H_0=M^2/(2E)$ and Hermitian $g$, as in Sec.~\ref{sec:redfield}.
The dot denotes the derivative of the field along the trajectory.
Set production at $s=0$ and detection at $s=T\simeq L$, with
$a_P=a(0)$ and $a_D=a(T)$.

The change $\psi_\nu=R\psi_\chi$, with $R=e^{\ii ag}$, gives
$H_\chi=R^\dagger H_0R$ in Eq.~\eqref{eq:representation}.
The evolution operators obey
\[
 S_\nu(D,P)=R_D S_\chi(D,P)R_P^\dagger.
\]
The end factors implement the change of variables at production and
detection. Expanding the transformed Hamiltonian gives
\begin{equation}
 H_\chi=H_0+\ii a[H_0,g]
 +\frac{a^2}{2}[g,[H_0,g]]+\cdots.
 \label{eq:commexp}
\end{equation}
The linear term is $a\mathcal J$, with
$\mathcal J=\ii[H_0,g]$. Through second order, the full evolution
contains the quadratic term, two insertions of $a\mathcal J$, and
the expansions of the end factors.

To check the relation directly, write $S_0(s)=e^{-\ii H_0s}$.
The derivative interaction gives
$\delta S_\nu^{(1)}=\ii\int_0^T\dot a(s)S_0(T-s)gS_0(s)\dd s$.
Since
\[
 \frac{\dd}{\dd s}\bigl[S_0(T-s)gS_0(s)\bigr]
 =\ii S_0(T-s)[H_0,g]S_0(s),
\]
integration by parts yields
\begin{equation}
 \begin{aligned}
 \delta S_\nu^{(1)}={}&\ii a_DgS_0(T)-\ii a_PS_0(T)g\\
 &+\int_0^T\!\dd s\,a(s)S_0(T-s)[H_0,g]S_0(s).
 \end{aligned}
 \label{eq:dysonIBP}
\end{equation}
The first two terms come from $R_D$ and $R_P^\dagger$; the
integral is generated by $a\mathcal J$ during propagation.
For diagonal $g$, all commutators vanish. The end factors then
recover Eq.~\eqref{eq:amplitude} exactly for any $ga$ within the
reduced Hamiltonian.

We now treat the full coupling matrix $g$ as a perturbation to the
vacuum evolution. For $i\ne j$, the first-order amplitude is
$\ii g_{ij}\int_0^T\dot a(s)e^{\ii\omega_{ij}s}\dd s$,
apart from a phase of unit modulus, where
$\omega_{ij}=\Delta m_{ij}^2/(2E)$.
Assume a real, zero-mean stationary field with correlation
$C_a(\tau)=\avg{a(s+\tau)a(s)}$, averaged over field realizations.
For a differentiable field with finite $\avg{\dot a^2}$ and a smooth
correlation, differentiating with respect to the two times gives
$\avg{\dot a(s+\tau)\dot a(s)}=-C_a''(\tau)$.

Squaring the amplitude gives the probability to second order in $g$:
\[
 \begin{aligned}
 \avg{P_{j\to i}(T)}\simeq{}&|g_{ij}|^2
 \int_{-T}^{T}\!\dd\tau\,(T-|\tau|)\\
 &\times e^{\ii\omega_{ij}\tau}[-C_a''(\tau)].
 \end{aligned}
\]
The factor $T-|\tau|$ counts the pairs of times separated by $\tau$
within the flight. When $T$ is much longer than the decay time of
the derivative correlation, replacing this factor by $T$ and extending
the integration limits gives a contribution $\Gamma_{j\to i}T$.
We use this perturbative result while the total transition probability
remains small.

Define the field power spectrum by
\[
 S_a(\omega)=\int_{-\infty}^{\infty}\!\dd\tau\,
 e^{\ii\omega\tau}C_a(\tau),
\]
and $S_{\dot a}$ by the same convention for the derivative
correlation. If $C_a$ and $C_a'$ vanish at large $|\tau|$ and
the integrals converge, two integrations by parts give
$S_{\dot a}(\omega)=\omega^2S_a(\omega)$. Thus
\begin{equation}
 \begin{aligned}
 \Gamma_{j\to i}
 &=|g_{ij}|^2S_{\dot a}(\omega_{ij})\\
 &=|\mathcal J_{ij}|^2S_a(\omega_{ij}),\qquad
 \omega_{ij}=\frac{\Delta m_{ij}^2}{2E}.
 \end{aligned}
 \label{eq:rate}
\end{equation}
Both representations give the same rate at order $g^2$. Under these
correlation assumptions, the end factors affect the probability over
a finite flight without changing its coefficient of linear growth.
The explicit $E^{-2}$ factor in $|\mathcal J_{ij}|^2$ is accompanied
by the spectrum evaluated at the energy-dependent frequency
$\omega_{ij}$.

\section{Finite exposure}
\label{app:exposure}

We calculate the interference factor for a finite observation of the
sinusoidal background in Sec.~\ref{sec:mode}. The field amplitude and
phase offset are held fixed during the observation.

For fixed source and detector positions, let $t$ be the production time
and write $\Delta\Phi_{ij}^{(a)}(t)=-A\sin(\omega_at+\beta)$.
Here $A=A_{ij}$ in Eq.~\eqref{eq:phaseamplitude} includes the field
variation during each flight. The constant $\beta$ absorbs the initial
field phase, the trajectory offset, and the overall sign. The
nonnegative weight $w(t)$ accounts for source intensity, detector live
time, and efficiency before applying the oscillation probability.
Averaging the interference term gives
\begin{equation}
 \mathcal C_{ij}^{\rm obs}=\int\dd t\,w(t)
 e^{\ii A\sin(\omega_at+\beta)},\qquad
 \int\dd t\,w(t)=1.
 \label{eq:exposure}
\end{equation}
Each reactor and energy contribution to Eq.~\eqref{eq:reactorspectrum}
uses its own time weight. The factor $\mathcal C_{ij}^{\rm obs}$ is
generally complex: its modulus
multiplies the oscillation contrast, while its argument shifts the
interference phase.

Using the Jacobi--Anger expansion~\cite[pp.~22--23]{Watson1944},
$e^{\ii A\sin\theta}=\sum_nJ_n(A)e^{\ii n\theta}$, we obtain
\begin{equation}
 \mathcal C_{ij}^{\rm obs}
 =\sum_{n=-\infty}^{\infty}J_n(A)e^{\ii n\beta}
 \int\dd t\,w(t)e^{\ii n\omega_at}.
 \label{eq:harmonics}
\end{equation}
Here $J_n$ is the Bessel function of the first kind of integer order $n$.
The time integral gives the weight of each harmonic $n\omega_a$ in
the observation.

For uniform exposure from $t_0$ to $t_0+T$, take $w(t)=1/T$ inside
the interval and zero outside it. Here $T$ is the observation time,
while each neutrino flight lasts approximately $L$.
With midpoint $t_c=t_0+T/2$ and $\tau=t-t_c$,
\[
 \begin{aligned}
 \frac{1}{T}\int_{t_0}^{t_0+T}\!\dd t\,e^{\ii n\omega_at}
 &=\frac{e^{\ii n\omega_at_c}}{T}
   \int_{-T/2}^{T/2}\!\dd\tau\,e^{\ii n\omega_a\tau}\\
 &=e^{\ii n\omega_at_c}
   \operatorname{sinc}\!\left(\frac{n\omega_aT}{2}\right),
 \end{aligned}
\]
where $\operatorname{sinc}z=\sin z/z$ and
$\operatorname{sinc}0=1$. Thus
\begin{equation}
 \mathcal C_{ij}^{(T)}=\sum_{n=-\infty}^{\infty}J_n(A)
 e^{\ii n(\omega_at_c+\beta)}
 \operatorname{sinc}\!\left(\frac{n\omega_aT}{2}\right).
 \label{eq:finiteaverage}
\end{equation}

For $\omega_aT\ll1$, expand the phase factor about $t_c$, with
$\Theta_c=\omega_at_c+\beta$. Odd powers of $\tau$ vanish in the
symmetric integral, and the average of $\tau^2$ is $T^2/12$.
At fixed $A$, this gives
\begin{equation}
 \begin{aligned}
 \mathcal C_{ij}^{(T)}=e^{\ii A\sin\Theta_c}
 \bigg[1-\frac{(\omega_aT)^2}{24}\big(
 A^2\cos^2\Theta_c\\
 {}+\ii A\sin\Theta_c\big)+\mathcal O((\omega_aT)^4)\bigg].
 \end{aligned}
 \label{eq:shortexposure}
\end{equation}
The modulus approaches one as $\omega_aT\to0$, since the events acquire
nearly the same additional phase. The leading decrease in the modulus
is half the leading phase variance; the imaginary correction shifts the
averaged phase relative to its midpoint value. The expansion requires
little variation of the induced phase across the observation, so its
accuracy depends on $A$ as well as on $\omega_aT$.

With uniform exposure over many periods, the sinc factors with $n\ne0$
tend to zero. At fixed $A$, the Bessel series is absolutely convergent,
so the limit can be taken term by term. The result approaches $J_0(A)$,
as in Eq.~\eqref{eq:bessel}. An integer number of periods gives this
result exactly, independently of the starting phase. For general
weights, the same result is approached when the time integrals in
Eq.~\eqref{eq:harmonics} are small for nonzero $n$ with appreciable
$J_n(A)$.

To describe uncertainty in the initial phase, one can average
Eq.~\eqref{eq:exposure} uniformly over $\beta\in[0,2\pi)$.
This gives $J_0(A)$ for any normalized $w(t)$, even for a short
observation. This average combines predictions for different possible initial
phases. In a single realization, the temporal average in
Eq.~\eqref{eq:exposure} combines the phases actually sampled during
the observation.

\section{Invisible decay rates and meson spectra}
\label{app:decays}

We calculate the Higgs and charged-meson decay rates used in
Sec.~\ref{sec:ew} at tree level, neglecting neutrino masses. The Higgs
rate uses the neutrino-selective dimension-seven operator at $m_h$,
with $m_a=0$. For mesons, we use the neutrino-only interaction and
the low-energy weak charged current, retaining the charged-lepton and
ALP masses.

The term linear in $h$ in Eq.~\eqref{eq:d7matching} gives
$h\to a\nu_i\bar\nu_j$. Let $p$, $q$, and $k$ be the neutrino,
antineutrino, and ALP momenta. For massless final particles, define
$s=2p\cdot k$ and $t=2q\cdot k$, the squared invariant masses of
the two neutrino--ALP pairs. The amplitude and its spin sum are
\begin{equation}
 \begin{aligned}
 \mathcal M_{ij}&=\frac{2g_{ij}}{v}
 k_\mu\bar u_i(p)\gamma^\mu P_Lv_j(q),\\
 \sum_{\rm spins}|\mathcal M_{ij}|^2
 &=\frac{4|g_{ij}|^2}{v^2}st.
 \end{aligned}
 \label{eq:htrace}
\end{equation}
Contracting the spin trace with the two ALP momenta gives
$4(p\cdot k)(q\cdot k)=st$. The scalar initial state requires no spin
average. The phase-space region is $s,t\geq0$, $s+t\leq m_h^2$, with
\[
 \dd\Gamma=\frac{\sum|\mathcal M_{ij}|^2}{256\pi^3m_h^3}\,\dd s\,\dd t.
\]
The remaining integral is
\[
 \int_0^{m_h^2}\!\dd s\int_0^{m_h^2-s}\!\dd t\,st
 =\frac{m_h^8}{24}.
\]
Combining this with Eq.~\eqref{eq:htrace} gives
$m_h^5|g_{ij}|^2/(1536\pi^3v^2)$ for each final-state pair.
Summing over the final pairs gives the factor
$\sum_{ij}|g_{ij}|^2=\Tr(g^\dagger g)$ and reproduces
Eq.~\eqref{eq:hwidth}.

For diagonal Majorana couplings, we use the same coefficient of the
canonically normalized left-handed Weyl current. The two assignments
of opposite helicities to the final momenta compensate the
identical-particle factor $1/2$, giving the same inclusive massless width.

At meson energies, the weak charged current reduces to a local
four-fermion interaction. The neutrino rotation in Sec.~\ref{sec:ibp}
adds a vertex that emits one ALP. For $P^-=\pi^-,K^-$ with momentum
$p_P$, we define the decay constant by
$\langle0|\bar q\gamma^\mu\gamma^5Q|P^-(p_P)\rangle
=\ii f_Pp_P^\mu$. With the quark-mixing element $V_{qQ}$ specified in
Sec.~\ref{sec:charged}, the amplitude is, up to an overall phase,
\begin{equation}
 \mathcal M(P^-\to\ell^-\bar\nu_i a)=
 \sqrt2G_FV_{qQ}f_P(Ug)_{\ell i}
 p_{P\mu}\bar u_\ell\gamma^\mu P_Lv_i.
 \label{eq:mesonamp}
\end{equation}
The same amplitude follows by attaching the ALP to the outgoing
neutrino line in the original derivative interaction. The external
massless-neutrino equation of motion cancels the propagator against
the momentum factor at the derivative vertex. Writing $p_\ell$ and
$p_\nu$ for the charged-lepton and antineutrino momenta, the spin sum is
\[
\begin{aligned}
 \sum_{\rm spins}|\mathcal M_i|^2
 ={}&4G_F^2|V_{qQ}|^2f_P^2|(Ug)_{\ell i}|^2\\
 &\times\bigl[2(p_P\!\cdot p_\ell)(p_P\!\cdot p_\nu)
       -m_P^2(p_\ell\!\cdot p_\nu)\bigr].
\end{aligned}
\]
The neutrino mass states are orthogonal, so their rates add, replacing
$|(Ug)_{\ell i}|^2$ by $G_\ell^2$ of Eq.~\eqref{eq:row}.
The charge-conjugate decay has the same width at this order.

For the phase-space integral, we use the invariant mass of the
neutrino--ALP pair, $q_{\rm miss}^2=(p_\nu+k)^2=(p_P-p_\ell)^2$.
Define $r=m_\ell^2/m_P^2$, $u=m_a^2/m_P^2$, and
$z=q_{\rm miss}^2/m_P^2$. The pair's mass ranges from $m_a$ to
$m_P-m_\ell$, so $u\leq z\leq(1-\sqrt r)^2$. The decay is open
for $\sqrt r+\sqrt u<1$ and its rate vanishes at and above threshold.
With $\lambda(1,r,z)=(1-r-z)^2-4rz$, the charged-lepton momentum
in the meson rest frame is $m_P\sqrt{\lambda(1,r,z)}/2$.

At fixed $z$, averaging the bracket in the spin sum over the neutrino
direction in the pair rest frame gives
\[
 \frac{m_P^4}{4}\left(1-\frac{u}{z}\right)
 [r-r^2+2rz+z-z^2].
\]
The angle-integrated phase-space measure is
$\dd\Phi_3=m_P^2\sqrt{\lambda(1,r,z)}(1-u/z)\dd z/(128\pi^3)$.
Combining it with the averaged spin sum and the factor $1/(2m_P)$
gives Eq.~\eqref{eq:meson}, with
\begin{equation}
 \begin{aligned}
 F(r,u)={}&12\!\int_u^{(1-\sqrt r)^2}\!\dd z\,
 \sqrt{\lambda(1,r,z)}\left(1-\frac{u}{z}\right)^2\\
 &\times[r-r^2+2rz+z-z^2].
 \end{aligned}
 \label{eq:Fmeson}
\end{equation}
The recoil supplies the factor $\sqrt{\lambda}$, while the squared
amplitude and the pair phase space each contribute one factor $1-u/z$.
The integrand, including the factor of 12, divided by $F(r,u)$ gives
the normalized distribution in $z$. A further factor of $1/m_P^2$
converts it to the distribution in $q_{\rm miss}^2$ plotted in
Fig.~\ref{fig:mesons}.

For a massless charged lepton, the integral reduces to
\begin{equation}
 F(0,u)=1-8u+8u^3-u^4-12u^2\ln u,
 \qquad F(0,0)=1.
 \label{eq:Fzero}
\end{equation}
This agrees with Ref.~\cite{AltmannshoferDrorGori2023}; the spectrum
with finite charged-lepton mass also agrees with the neutrino-only
term in Eq.~(4.6) of Ref.~\cite{ExoticPions2026}, after matching the
current normalization.

We evaluate the branching ratios as
$\mathrm{BR}(P\to\ell\nu a)=\tau_P\Gamma(P\to\ell\nu a)$,
with the lifetime $\tau_P$ expressed in natural units. For pions we use
$m_\pi=139.57039$ MeV, $f_\pi=130.2$ MeV,
$\tau_\pi=2.6033\times10^{-8}$ s, and $|V_{ud}|=0.97373$;
for kaons we use $m_K=493.677$ MeV, $f_K=155.7$ MeV,
$\tau_K=1.238\times10^{-8}$ s, and $|V_{us}|=0.2243$.
The remaining inputs are $m_e=0.510999$ MeV,
$m_\mu=105.6584$ MeV, and $G_F=1.1663787\times10^{-5}\GeV^{-2}$.
At $m_a=0$, the integral gives $F(m_\mu^2/m_\pi^2,0)=0.05914$ and
$F(m_\mu^2/m_K^2,0)=0.9460$. These inputs reproduce the branching-ratio
coefficients in Sec.~\ref{sec:charged}.

For the numerical flavor comparisons in Sec.~\ref{sec:ew}, we use
$\sin^2\theta_{12}=0.307$, $\sin^2\theta_{13}=0.022$,
$\sin^2\theta_{23}=0.55$, and $\delta_{\rm CP}=1.2\pi$.
These illustrative values lie within the individual normal-ordering
ranges of Ref.~\cite{NuFIT2024}. In the usual mixing convention,
\[
 \begin{aligned}
 U_{e1}&=\cos\theta_{12}\cos\theta_{13},\\
 U_{\mu1}&=-\sin\theta_{12}\cos\theta_{23}\\
 &\quad-\cos\theta_{12}\sin\theta_{23}\sin\theta_{13}
 e^{\ii\delta_{\rm CP}}.
 \end{aligned}
\]
This gives $|U_{e1}|^2\simeq0.678$ and
$|U_{\mu1}|^2\simeq0.0915$. For the muon-decay comparison in
Sec.~\ref{sec:d5}, the required product is
$|U_{e1}U_{\mu1}^*|\simeq0.249$. We use the same $m_\mu$ and
$\tau_\mu=2.1969811\times10^{-6}$ s, with $\Gamma_\mu=1/\tau_\mu$
in natural units.

For the NA62 comparison, the measured $K^+\to\mu^+\nu$ sample
gives an effective kaon normalization
$N_K=(1.14\pm0.02)\times10^{10}$~\cite{NA622021}. It already
includes trigger losses and vetoes from unrelated detector activity
shared by the signal and normalization channels.

Let $\Pi$ denote the decay kinematics and $p_a(\Pi)$ the normalized
$K\to\mu\nu a$ signal distribution, with
$\int\dd\Pi\,p_a(\Pi)=1$. In this normalization, $\epsilon(\Pi)$
is the probability of passing the remaining selections, averaged over
the beam and decay positions. It includes detector resolution and cuts
on the measured momenta. The acceptance $\mathcal A_a$ and expected
selected signal count $N_s$ satisfy
\begin{equation}
 N_s=N_K\,\mathrm{BR}(K\to\mu\nu a)\,\mathcal A_a,
 \quad
 \mathcal A_a=\int\dd\Pi\,p_a(\Pi)\epsilon(\Pi).
 \label{eq:acceptance}
\end{equation}

NA62 observes 6894 events with reconstructed
$q_{\rm miss}^2>0.1\GeV^2$, for an expected background of
$7549\pm928$ events. Its counting analysis gives an upper limit of
1184 selected signal events at 90\% confidence. The published
$K^+\to\mu^+\nu\bar\nu\nu$ branching-ratio limit uses the
acceptance of its four-body spectrum. For our spectrum, we use the
event limit with $\mathcal A_a$ in Eq.~\eqref{eq:acceptance},
retaining the same region and background treatment.

For $m_a=0$, the fraction of our spectrum above the same threshold,
before detector response, is $f_{\rm cut}=0.26556$. Neglecting shifts
across the threshold caused by momentum resolution, we write
$\mathcal A_a\simeq f_{\rm cut}\bar\epsilon$, where
$\bar\epsilon$ is the average remaining selection probability for
signal events in that interval. Using the central $N_K$ and the
calculated branching ratio gives
\[
 \begin{aligned}
 G_\mu&<\left[
 \frac{1184}{N_K(9.09\times10^{-4}\GeV^2)f_{\rm cut}\bar\epsilon}
 \right]^{1/2}\\
 &=\frac{0.02074}{\sqrt{\bar\epsilon}}\GeV^{-1}.
 \end{aligned}
\]

Oscillations depend on coupling differences, so we minimize $G_\mu^2$
over the common part of the $g_i$. For $g_2=g_3$, write $g_1=c$ and
$g_2=g_3=c+\Delta g$, with $\Delta g=\Delta g_{21}=\Delta g_{31}$.
Unitarity gives $\sum_{i=2,3}|U_{\mu i}|^2=1-|U_{\mu1}|^2$.
Writing $w_\mu=|U_{\mu1}|^2$ and completing the square yields
\[
 \begin{aligned}
 G_\mu^2&=w_\mu c^2+(1-w_\mu)(c+\Delta g)^2\\
 &=[c+(1-w_\mu)\Delta g]^2\\
 &\quad+w_\mu(1-w_\mu)(\Delta g)^2\\
 &\geq w_\mu(1-w_\mu)(\Delta g)^2.
 \end{aligned}
\]
The minimum occurs at $c=-(1-w_\mu)\Delta g$. Combining it with
the limit on $G_\mu$ and the mixing inputs above gives the conditional
relation $|\Delta g|\lesssim0.072/\sqrt{\bar\epsilon}\GeV^{-1}$.
The signal acceptance and its uncertainty remain to be evaluated.

\bibliography{references}

@article{ATLASInvisible2023,
  author        = {{ATLAS Collaboration}},
  title         = {{Combination of searches for invisible decays of the Higgs boson using 139 fb$^{-1}$ of proton-proton collision data at $\sqrt{s}=13$ TeV collected with the ATLAS experiment}},
  journal       = {Phys. Lett. B},
  volume        = {842},
  pages         = {137963},
  year          = {2023},
  doi           = {10.1016/j.physletb.2023.137963},
  eprint        = {2301.10731},
  archivePrefix = {arXiv},
  primaryClass  = {hep-ex}
}

@article{AdsheadLozanov2022,
  author        = {Adshead, Peter and Lozanov, Kaloian D.},
  title         = {{Axion anomalies}},
  journal       = {{JHEP}},
  volume        = {08},
  pages         = {077},
  year          = {2022},
  doi           = {10.1007/JHEP08(2022)077},
  eprint        = {2112.07645},
  archivePrefix = {arXiv},
  primaryClass  = {hep-th}
}

@article{Airoldi2026,
  author        = {Airoldi, Lua F. T. and Alves, Gustavo F. S. and Machado, Pedro A. N. and Vander Griend, Peter},
  title         = {{Open system approach to neutrinos propagating in an ultralight scalar background}},
  journal       = {Phys. Rev. D},
  volume        = {114},
  number        = {1},
  pages         = {015043},
  year          = {2026},
  doi           = {10.1103/rh5j-fvry},
  eprint        = {2603.02382},
  archivePrefix = {arXiv},
  primaryClass  = {hep-ph}
}

@article{AltmannshoferDrorGori2023,
  author        = {Altmannshofer, Wolfgang and Dror, Jeff A. and Gori, Stefania},
  title         = {{New Opportunities for Detecting Axion-Lepton Interactions}},
  journal       = {Phys. Rev. Lett.},
  volume        = {130},
  number        = {24},
  pages         = {241801},
  year          = {2023},
  doi           = {10.1103/PhysRevLett.130.241801},
  eprint        = {2209.00665},
  archivePrefix = {arXiv},
  primaryClass  = {hep-ph}
}

@article{Arias2012,
  author        = {Arias, Paola and Cadamuro, Davide and Goodsell, Mark and Jaeckel, Joerg and Redondo, Javier and Ringwald, Andreas},
  title         = {{WISPy} cold dark matter},
  journal       = {{JCAP}},
  volume        = {06},
  pages         = {013},
  year          = {2012},
  doi           = {10.1088/1475-7516/2012/06/013},
  eprint        = {1201.5902},
  archivePrefix = {arXiv},
  primaryClass  = {hep-ph}
}

@article{BenattiFloreanini2000,
  author        = {Benatti, Fabio and Floreanini, Roberto},
  title         = {{Open system approach to neutrino oscillations}},
  journal       = {{JHEP}},
  volume        = {02},
  pages         = {032},
  year          = {2000},
  doi           = {10.1088/1126-6708/2000/02/032},
  eprint        = {hep-ph/0002221},
  archivePrefix = {arXiv},
  primaryClass  = {hep-ph}
}

@article{Berlin2016,
  author        = {Berlin, Asher},
  title         = {{Neutrino Oscillations as a Probe of Light Scalar Dark Matter}},
  journal       = {Phys. Rev. Lett.},
  volume        = {117},
  number        = {23},
  pages         = {231801},
  year          = {2016},
  doi           = {10.1103/PhysRevLett.117.231801},
  eprint        = {1608.01307},
  archivePrefix = {arXiv},
  primaryClass  = {hep-ph}
}

@article{Berlinetal2024,
  author        = {Berlin, Asher and Millar, Alexander J. and Trickle, Tanner and Zhou, Kevin},
  title         = {Physical signatures of fermion-coupled axion dark matter},
  journal       = {{JHEP}},
  volume        = {05},
  pages         = {314},
  year          = {2024},
  doi           = {10.1007/JHEP05(2024)314},
  eprint        = {2312.11601},
  archivePrefix = {arXiv},
  primaryClass  = {hep-ph},
  issn          = {1029-8479}
}

@article{BonillaEtAl2024,
  author        = {Bonilla, J. and Gavela, B. and Machado-Rodr\'iguez, J.},
  title         = {{Limits on ALP-neutrino couplings from loop-level processes}},
  journal       = {Phys. Rev. D},
  volume        = {109},
  number        = {5},
  pages         = {055023},
  year          = {2024},
  doi           = {10.1103/PhysRevD.109.055023},
  eprint        = {2309.15910},
  archivePrefix = {arXiv},
  primaryClass  = {hep-ph}
}

@article{BrdarEtAl2018,
  author        = {Brdar, Vedran and Kopp, Joachim and Liu, Jia and Prass, Pascal and Wang, Xiao-Ping},
  title         = {{Fuzzy dark matter and nonstandard neutrino interactions}},
  journal       = {Phys. Rev. D},
  volume        = {97},
  number        = {4},
  pages         = {043001},
  year          = {2018},
  doi           = {10.1103/PhysRevD.97.043001},
  eprint        = {1705.09455},
  archivePrefix = {arXiv},
  primaryClass  = {hep-ph}
}

@article{CalibbiEtAl2021,
  author        = {Calibbi, Lorenzo and Redigolo, Diego and Ziegler, Robert and Zupan, Jure},
  title         = {{Looking forward to lepton-flavor-violating ALPs}},
  journal       = {{JHEP}},
  volume        = {09},
  pages         = {173},
  year          = {2021},
  doi           = {10.1007/JHEP09(2021)173},
  eprint        = {2006.04795},
  archivePrefix = {arXiv},
  primaryClass  = {hep-ph}
}

@article{Centers2021,
  author        = {Centers, Gary P. and Blanchard, John W. and Conrad, Jan and Figueroa, Nataniel L. and Garcon, Antoine and Gramolin, Alexander V. and Jackson Kimball, Derek F. and Lawson, Matthew and Pelssers, Bart and Smiga, Joseph A. and Sushkov, Alexander O. and Wickenbrock, Arne and Budker, Dmitry and Derevianko, Andrei},
  title         = {{Stochastic fluctuations of bosonic dark matter}},
  journal       = {Nature Commun.},
  volume        = {12},
  number        = {1},
  pages         = {7321},
  year          = {2021},
  doi           = {10.1038/s41467-021-27632-7},
  eprint        = {1905.13650},
  archivePrefix = {arXiv},
  primaryClass  = {astro-ph.CO}
}

@misc{Chathirathas2026,
  author        = {Chathirathas, Kierthika and Chatterjee, Sabya Sachi and Schwetz, Thomas},
  title         = {{Oscillating Neutrinos vs. Oscillating Scalars: Constraining Scalar Dark Matter-Induced Neutrino Mass}},
  year          = {2026},
  eprint        = {2608.06460},
  archivePrefix = {arXiv},
  primaryClass  = {hep-ph}
}

@article{Coloma2018,
  author        = {Coloma, Pilar and Lopez-Pavon, Jacobo and Martinez-Soler, Ivan and Nunokawa, Hiroshi},
  title         = {{Decoherence in neutrino propagation through matter, and bounds from IceCube/DeepCore}},
  journal       = {Eur. Phys. J. C},
  volume        = {78},
  number        = {8},
  pages         = {614},
  year          = {2018},
  doi           = {10.1140/epjc/s10052-018-6092-6},
  eprint        = {1803.04438},
  archivePrefix = {arXiv},
  primaryClass  = {hep-ph}
}

@article{CriadoPerezVictoria2019,
  author        = {Criado, J. C. and P{\'e}rez-Victoria, M.},
  title         = {{Field redefinitions in effective theories at higher orders}},
  journal       = {{JHEP}},
  volume        = {03},
  pages         = {038},
  year          = {2019},
  doi           = {10.1007/JHEP03(2019)038},
  eprint        = {1811.09413},
  archivePrefix = {arXiv},
  primaryClass  = {hep-ph}
}

@article{DeRomeri2023,
  author        = {De Romeri, Valentina and Giunti, Carlo and Stuttard, Thomas and Ternes, Christoph A.},
  title         = {{Neutrino oscillation bounds on quantum decoherence}},
  journal       = {{JHEP}},
  volume        = {09},
  pages         = {097},
  year          = {2023},
  doi           = {10.1007/JHEP09(2023)097},
  eprint        = {2306.14699},
  archivePrefix = {arXiv},
  primaryClass  = {hep-ph}
}

@article{DevMachadoMartinez2021,
  author        = {Dev, Abhish and Machado, Pedro A. N. and Mart\'inez-Mirav\'e, Pablo},
  title         = {{Signatures of ultralight dark matter in neutrino oscillation experiments}},
  journal       = {{JHEP}},
  volume        = {01},
  pages         = {094},
  year          = {2021},
  doi           = {10.1007/JHEP01(2021)094},
  eprint        = {2007.03590},
  archivePrefix = {arXiv},
  primaryClass  = {hep-ph}
}

@article{DreinerHaberMartin2010,
  author        = {Dreiner, Herbi K. and Haber, Howard E. and Martin, Stephen P.},
  title         = {{Two-component spinor techniques and Feynman rules for quantum field theory and supersymmetry}},
  journal       = {Phys. Rept.},
  volume        = {494},
  number        = {1--2},
  pages         = {1--196},
  year          = {2010},
  doi           = {10.1016/j.physrep.2010.05.002},
  eprint        = {0812.1594},
  archivePrefix = {arXiv},
  primaryClass  = {hep-ph}
}

@misc{ExoticPions2026,
  author        = {Altmannshofer, Wolfgang and Dror, Jeff A. and Giffin, Pierce and Gori, Stefania and Jackson, Ollie and Luong, Khai and Schwendimann, Patrick and Seo, Se Rang},
  title         = {{Exploring Invisible New Physics with Exotic Pion Decays}},
  year          = {2026},
  eprint        = {2601.06254},
  archivePrefix = {arXiv},
  primaryClass  = {hep-ph}
}

@article{ForeroParkeTernesFunchal2021,
  author        = {Forero, David V. and Parke, Stephen J. and Ternes, Christoph A. and Zukanovich Funchal, Renata},
  title         = {{JUNO's prospects for determining the neutrino mass ordering}},
  journal       = {Phys. Rev. D},
  volume        = {104},
  number        = {11},
  pages         = {113004},
  year          = {2021},
  doi           = {10.1103/PhysRevD.104.113004},
  eprint        = {2107.12410},
  archivePrefix = {arXiv},
  primaryClass  = {hep-ph}
}

@article{Giannottietal2017,
  author        = {Giannotti, Maurizio and Irastorza, Igor G. and Redondo, Javier and Ringwald, Andreas and Saikawa, Ken'ichi},
  title         = {Stellar recipes for axion hunters},
  journal       = {{JCAP}},
  volume        = {10},
  pages         = {010},
  year          = {2017},
  doi           = {10.1088/1475-7516/2017/10/010},
  eprint        = {1708.02111},
  archivePrefix = {arXiv},
  primaryClass  = {hep-ph}
}

@article{GrahamRajendran2013,
  author        = {Graham, Peter W. and Rajendran, Surjeet},
  title         = {{New observables for direct detection of axion dark matter}},
  journal       = {Phys. Rev. D},
  volume        = {88},
  number        = {3},
  pages         = {035023},
  year          = {2013},
  doi           = {10.1103/PhysRevD.88.035023},
  eprint        = {1306.6088},
  archivePrefix = {arXiv},
  primaryClass  = {hep-ph}
}

@article{Gramolin2022,
  author        = {Gramolin, Alexander V. and Wickenbrock, Arne and Aybas, Deniz and Bekker, Hendrik and Budker, Dmitry and Centers, Gary P. and Figueroa, Nataniel L. and Jackson Kimball, Derek F. and Sushkov, Alexander O.},
  title         = {{Spectral signatures of axionlike dark matter}},
  journal       = {Phys. Rev. D},
  volume        = {105},
  number        = {3},
  pages         = {035029},
  year          = {2022},
  doi           = {10.1103/PhysRevD.105.035029},
  eprint        = {2107.11948},
  archivePrefix = {arXiv},
  primaryClass  = {hep-ph}
}

@article{GrojeanKleyYao2023,
  author        = {Grojean, Christophe and Kley, Jonathan and Yao, Chang-Yuan},
  title         = {{Hilbert series for ALP EFTs}},
  journal       = {{JHEP}},
  volume        = {11},
  pages         = {196},
  year          = {2023},
  doi           = {10.1007/JHEP11(2023)196},
  eprint        = {2307.08563},
  archivePrefix = {arXiv},
  primaryClass  = {hep-ph}
}

@techreport{HiggsWG,
  author        = {Heinemeyer, S. and others},
  collaboration = {LHC Higgs Cross Section Working Group},
  title         = {{Handbook of LHC Higgs Cross Sections: 3. Higgs Properties}},
  institution   = {CERN},
  number        = {CERN-2013-004},
  year          = {2013},
  doi           = {10.5170/CERN-2013-004},
  eprint        = {1307.1347},
  archivePrefix = {arXiv},
  primaryClass  = {hep-ph}
}

@article{HuangNath2018,
  author        = {Huang, Guo-yuan and Nath, Newton},
  title         = {{Neutrinophilic axion-like dark matter}},
  journal       = {Eur. Phys. J. C},
  volume        = {78},
  number        = {11},
  pages         = {922},
  year          = {2018},
  doi           = {10.1140/epjc/s10052-018-6391-y},
  eprint        = {1809.01111},
  archivePrefix = {arXiv},
  primaryClass  = {hep-ph}
}

@article{Hui2021,
  author        = {Hui, Lam},
  title         = {{Wave Dark Matter}},
  journal       = {Ann. Rev. Astron. Astrophys.},
  volume        = {59},
  number        = {1},
  pages         = {247--289},
  year          = {2021},
  doi           = {10.1146/annurev-astro-120920-010024},
  eprint        = {2101.11735},
  archivePrefix = {arXiv},
  primaryClass  = {astro-ph.CO}
}

@article{IceCube2024,
  author        = {{IceCube Collaboration}},
  title         = {{Search for decoherence from quantum gravity with atmospheric neutrinos}},
  journal       = {Nature Phys.},
  volume        = {20},
  number        = {6},
  pages         = {913--920},
  year          = {2024},
  doi           = {10.1038/s41567-024-02436-w},
  eprint        = {2308.00105},
  archivePrefix = {arXiv},
  primaryClass  = {hep-ex}
}

@article{JUNOPhysics2022,
  author        = {{JUNO Collaboration}},
  title         = {{JUNO physics and detector}},
  journal       = {Prog. Part. Nucl. Phys.},
  volume        = {123},
  pages         = {103927},
  year          = {2022},
  doi           = {10.1016/j.ppnp.2021.103927},
  eprint        = {2104.02565},
  archivePrefix = {arXiv},
  primaryClass  = {hep-ex}
}

@article{KamLAND2008,
  author        = {Abe, S. and others},
  collaboration = {KamLAND},
  title         = {{Precision Measurement of Neutrino Oscillation Parameters with KamLAND}},
  journal       = {Phys. Rev. Lett.},
  volume        = {100},
  number        = {22},
  pages         = {221803},
  year          = {2008},
  doi           = {10.1103/PhysRevLett.100.221803},
  eprint        = {0801.4589},
  archivePrefix = {arXiv},
  primaryClass  = {hep-ex}
}

@article{KamLAND2013,
  author        = {Gando, A. and others},
  collaboration = {KamLAND},
  title         = {{Reactor on-off antineutrino measurement with KamLAND}},
  journal       = {Phys. Rev. D},
  volume        = {88},
  number        = {3},
  pages         = {033001},
  year          = {2013},
  doi           = {10.1103/PhysRevD.88.033001},
  eprint        = {1303.4667},
  archivePrefix = {arXiv},
  primaryClass  = {hep-ex}
}

@article{Lichkunov2025,
  author        = {Lichkunov, Alexey and Stankevich, Konstantin and Studenikin, Alexander},
  title         = {{Neutrino quantum decoherence in a fluctuating ALPs field}},
  journal       = {Phys. Rev. D},
  volume        = {112},
  number        = {12},
  pages         = {123007},
  year          = {2025},
  doi           = {10.1103/16ml-srzs},
  eprint        = {2509.22554},
  archivePrefix = {arXiv},
  primaryClass  = {hep-ph}
}

@article{LosadaEtAl2022,
  author        = {Losada, Marta and Nir, Yosef and Perez, Gilad and Shpilman, Yogev},
  title         = {{Probing scalar dark matter oscillations with neutrino oscillations}},
  journal       = {{JHEP}},
  volume        = {04},
  pages         = {030},
  year          = {2022},
  doi           = {10.1007/JHEP04(2022)030},
  eprint        = {2107.10865},
  archivePrefix = {arXiv},
  primaryClass  = {hep-ph}
}

@article{NA622021,
  author        = {Cortina Gil, E. and others},
  collaboration = {NA62},
  title         = {{Search for $K^+$ decays to a muon and invisible particles}},
  journal       = {Phys. Lett. B},
  volume        = {816},
  pages         = {136259},
  year          = {2021},
  doi           = {10.1016/j.physletb.2021.136259},
  eprint        = {2101.12304},
  archivePrefix = {arXiv},
  primaryClass  = {hep-ex}
}

@article{NuFIT2024,
  author        = {Esteban, Ivan and Gonzalez-Garcia, M. C. and Maltoni, Michele and Martinez-Soler, Ivan and Pinheiro, Jo{\~a}o Paulo and Schwetz, Thomas},
  title         = {{NuFit-6.0: updated global analysis of three-flavor neutrino oscillations}},
  journal       = {{JHEP}},
  volume        = {12},
  pages         = {216},
  year          = {2024},
  doi           = {10.1007/JHEP12(2024)216},
  eprint        = {2410.05380},
  archivePrefix = {arXiv},
  primaryClass  = {hep-ph}
}

@article{ORCA2025,
  author        = {{KM3NeT Collaboration}},
  title         = {{Search for quantum decoherence in neutrino oscillations with six detection units of KM3NeT/ORCA}},
  journal       = {{JCAP}},
  volume        = {03},
  pages         = {039},
  year          = {2025},
  doi           = {10.1088/1475-7516/2025/03/039},
  eprint        = {2410.01388},
  archivePrefix = {arXiv},
  primaryClass  = {hep-ex}
}

@article{PIENU2021,
  author        = {Aguilar-Arevalo, A. and others},
  collaboration = {PIENU},
  title         = {{Search for three body pion decays $\pi^+\to l^+\nu X$}},
  journal       = {Phys. Rev. D},
  volume        = {103},
  number        = {5},
  pages         = {052006},
  year          = {2021},
  doi           = {10.1103/PhysRevD.103.052006},
  eprint        = {2101.07381},
  archivePrefix = {arXiv},
  primaryClass  = {hep-ex}
}

@article{RogersPeiris2021,
  author        = {Rogers, Keir K. and Peiris, Hiranya V.},
  title         = {{Strong Bound on Canonical Ultralight Axion Dark Matter from the Lyman-Alpha Forest}},
  journal       = {Phys. Rev. Lett.},
  volume        = {126},
  number        = {7},
  pages         = {071302},
  year          = {2021},
  doi           = {10.1103/PhysRevLett.126.071302},
  eprint        = {2007.12705},
  archivePrefix = {arXiv},
  primaryClass  = {astro-ph.CO}
}

@book{Schlosshauer2007,
  author        = {Schlosshauer, Maximilian},
  title         = {{Decoherence and the Quantum-to-Classical Transition}},
  publisher     = {Springer},
  address       = {Berlin, Heidelberg},
  series        = {The Frontiers Collection},
  year          = {2007},
  doi           = {10.1007/978-3-540-35775-9},
  isbn          = {978-3-540-35773-5}
}

@article{TWIST2015,
  author        = {Bayes, R. and others},
  collaboration = {TWIST},
  title         = {{Search for two body muon decay signals}},
  journal       = {Phys. Rev. D},
  volume        = {91},
  number        = {5},
  pages         = {052020},
  year          = {2015},
  doi           = {10.1103/PhysRevD.91.052020},
  eprint        = {1409.0638},
  archivePrefix = {arXiv},
  primaryClass  = {hep-ex}
}

@article{TernesSN2025,
  author        = {Ternes, Christoph A. and Pagliaroli, Giulia and Villante, Francesco L.},
  title         = {{SN1987A bounds on neutrino quantum decoherence}},
  journal       = {Phys. Rev. D},
  volume        = {112},
  number        = {6},
  pages         = {063058},
  year          = {2025},
  doi           = {10.1103/4j6r-d2wx},
  eprint        = {2503.04573},
  archivePrefix = {arXiv},
  primaryClass  = {hep-ph}
}

@book{Watson1944,
  author        = {Watson, G. N.},
  title         = {{A Treatise on the Theory of Bessel Functions}},
  publisher     = {Cambridge University Press},
  address       = {Cambridge},
  edition       = {2},
  year          = {1944}
}

@article{Winch2025,
  author        = {Winch, Harrison and Rogers, Keir K. and Hlo{\v{z}}ek, Ren{\'e}e and Marsh, David J. E.},
  title         = {{High-redshift, Small-scale Tests of Ultralight Axion Dark Matter Using Hubble and Webb Galaxy UV Luminosities}},
  journal       = {Astrophys. J.},
  volume        = {976},
  number        = {1},
  pages         = {40},
  year          = {2024},
  doi           = {10.3847/1538-4357/ad7a73},
  eprint        = {2404.11071},
  archivePrefix = {arXiv},
  primaryClass  = {astro-ph.CO}
}

@article{deSalasWidmark2021,
  author        = {de Salas, Pablo F. and Widmark, Axel},
  title         = {{Dark matter local density determination: recent observations and future prospects}},
  journal       = {Rept. Prog. Phys.},
  volume        = {84},
  number        = {10},
  pages         = {104901},
  year          = {2021},
  doi           = {10.1088/1361-6633/ac24e7},
  eprint        = {2012.11477},
  archivePrefix = {arXiv},
  primaryClass  = {astro-ph.GA}
}

@misc{BeccariaTernes2026,
  author        = {Beccaria, Martina and Ternes, Christoph A.},
  title         = {{Probing damping effects in neutrino oscillations with 207 days of JUNO data}},
  year          = {2026},
  eprint        = {2606.13362},
  archivePrefix = {arXiv},
  primaryClass  = {hep-ph},
  url           = {https://arxiv.org/abs/2606.13362v2},
  note          = {Version 2, 3 November 2026}
}

@misc{JUNO207Days2026,
  author        = {{JUNO Collaboration}},
  title         = {{Measurement of Neutrino Oscillation Parameters at JUNO and Indication of the Neutrino Mass Ordering}},
  year          = {2026},
  eprint        = {2609.37895},
  archivePrefix = {arXiv},
  primaryClass  = {hep-ex}
}
\end{document}